\documentclass[aps,prb,twocolumn,showpacs,floatfix]{revtex4}
\pdfoutput=1
\usepackage{graphicx}    
\usepackage{hyperref}    
\usepackage{bm}          
\usepackage[sumlimits,intlimits]{amsmath}
\usepackage{amsfonts,amssymb,cases}
\DeclareMathOperator{\sgn}{sgn}

\begin{document}

\title{Numerical studies of variable-range hopping in one-dimensional systems}

\author{A. S. Rodin}
\author{M. M. Fogler}
\affiliation{University of California San Diego, 9500 Gilman Drive, La
Jolla, California 92093}

\date{\today}
\begin{abstract}

Hopping transport in a one-dimensional system is studied numerically. A fast
algorithm is devised to find the lowest-resistance path at arbitrary electric field. Probability distribution functions of individual resistances on the path and the net resistance are calculated and fitted to compact analytic formulas. Qualitative differences between statistics of resistance fluctuations in Ohmic and non-Ohmic regimes are elucidated. The results are compared with prior theoretical and experimental work on the subject.

\end{abstract}


\maketitle


\section{Introduction}
\label{sec:Introduction}

It is well known that low-temperature transport in disordered one-dimensional (1D) structures is distinguished by large mesoscopic fluctuations. Such fluctuations have been measured~\cite{Fowler1982cir, Webb1986hci, Ladieu1993csiII, Aleshin2005cbt, Gao2006cao} even in samples of considerable length. They arise from the interplay of localization and rigid geometrical constraints on possible current paths. The total resistance tends to be dominated by a few strong obstacles --- ``breaks'' --- which occur at random due to disorder in the sample.~\cite{Kurkijarvi1973hci, Brenig1973hci, Shante1973hci, Lee1984vrh, Serota1986nao, Raikh1989fot} This unusual behavior can be contrasted with a more familiar case of dimensions $d > 1$. There the current can go around the breaks, so that the mesoscopic fluctuations of transport properties are usually small and self-averaging.

In this paper we consider 1D systems that are not too short, so that the coherent tunneling of electrons through their entire length~\cite{Azbel1984ftc, Tartakovski1993cfa} is extremely improbable. Instead, electrons traverse each sample via a sequence of many incoherent tunneling acts --- the variable-range hopping~\cite{Lee1984vrh} (VRH). By studying the VRH transport~\cite{Shklovskii1984epo} one aims to extract information about the nature of electron localization and disorder in the system. However, this task is far from trivial. Although the basic physics of the 1D VRH problem is quite well understood, experimental studies of VRH are typically done in a narrow parameter range where usual theoretical approximations are still rather crude. Below we demonstrate that large corrections appear when the transport properties of a standard VRH model are calculated numerically, which means, with fewer approximations.

To deal with large mesoscopic fluctuations we follow prior work and compute both the probability distribution functions~\cite{Serota1986nao,  Raikh1989fot, He2003spa, Ortuno2004cfi} (PDF) and suitable averages of the transport observables. For example, we study the ensemble-averaged conductance $\langle G \rangle$, which can be measured experimentally by having a large number of 1D wires connected in parallel.~\cite{Khavin1998slo}

Our primary purpose is to investigate non-Ohmic effects, e.g., the dependence of function $\langle G \rangle (F, T)$ on the electric force $F = -e E$. This regime has been studied much less compared to the Ohmic one. However, recently an analytical theory of non-Ohmic 1D VRH has been proposed in a work of one of us.~\cite{Fogler2005nov} Here we approach the same problem numerically. We have developed an efficient computer algorithm, which is able to find the VRH conductance of a given sample at arbitrary electric field. By choosing a low $F$ the Ohmic conductance $G(0, T)$ can also be calculated.

Since the Ohmic case has been more widely studied, it deserves a brief discussion first. In dimensions $d > 1$ the Ohmic conductance is known to follow the stretched exponential temperature dependence:
\begin{equation}\label{eqn:VRH}
 G(0, T) = G_0  \exp \left[- (\Delta / T)^\gamma \right]\,,
\end{equation}
where $\Delta$ is some energy scale, and $G_0$ depends on $T$ at most
algebraically. The exponent $\gamma = 1 / (d + 1)$ at $d > 1$ signifies the Mott law. The Efros-Shklovskii law corresponds to $\gamma = 1 / 2$. It applies when the long-range Coulomb interactions
are important.~\cite{Shklovskii1984epo}

In 1D, the Mott and Efros-Shklovskii exponents coincide. This is because in 1D the $1/r$ Coulomb potential is only marginally long-range to begin with, and then typically also screened by a nearby metallic gate. The importance of the remaining interactions is determined by the dimensionless parameter
\begin{equation}
\epsilon = 1 + (e^2 g / C)\,,
\label{eqn:epsilon}
\end{equation}
which has the physical meaning of the dielectric constant.  Here $C$ is the capacitance to the gate per unit length of the wire and $g$ is the average density of states. (Note that $\epsilon$ is related to the Luttinger-liquid parameter~\cite{Giamarchi2004qpi} of a disorder-free 1D system.) In this paper we study the case of weak interactions, $\epsilon \simeq 1$, where, naively, the Mott law may seem to be a reasonable starting point.

Actually, the 1D Mott law is modified by the aforementioned mesoscopic fluctuations. Lee~\cite{Lee1984vrh} and Raikh and Ruzin~\cite{Raikh1989fot} showed analytically that at low temperatures the energy scale $\Delta$ in Eq.~\eqref{eqn:VRH} is not a constant but a logarithmic function of $T$. More importantly, $\Delta(T)$ is determined not only by intrinsic properties of the system but also by its size. As $T$ increases, a narrow range of temperature appears where another dependence, $\Delta(T) \propto 1 / T$ is realized. Hence, instead of the $\gamma = 1 /2$ Mott law we effectively have a simple activation,~\cite{Kurkijarvi1973hci, Brenig1973hci} $\gamma = 1$. Such behavior has been confirmed by numerical simulations,~\cite{Lee1984vrh, Serota1986nao, Ortuno2004cfi, He2003spa, Deroulers2007dot} so it is
considered well established. 

Nevertheless, to establish a reference point for our study of non-Ohmic VRH we examined the \emph{Ohmic} conductance carefully by our method. Remarkably, we found that it is essential to introduce often discarded ``subleading'' terms in the analytical expressions. If this is not done, analytical and numerical results for $G$ can differ by orders of magnitude.

In the non-Ohmic regime, which is our main subject of interest,
it has been customary~\cite{Mott1971cin, Hill1971hci, Pollak1976apt, Larkin1982aci} to characterize the field-dependence of the conductivity by means of the length parameter $L_c$:
\begin{equation}\label{eqn:exponential}
 \langle G\rangle (F, T) = \langle G \rangle (0, T) \exp
 \left(\, |F| L_c /\, T\,	\right)\,.
\end{equation}
In experiment, this law typically  describes the first decade of
the conductivity rise. Thereafter, deviations tend to occur. Indeed, in theory~\cite{Shklovskii1976nhc, Levin1984lth, Talamantes1987mfv} $L_c$ is expected to be not a constant but a function of $F$ and $T$. We will show that in 1D $L_c$ may also depend on the averaging procedure utilized to obtain $\langle G \rangle$.

At large enough $F$, Eq.~\eqref{eqn:exponential} eventually becomes a poor approximation. Theoretically, it should cross over to~\cite{Fogler2005nov}
\begin{equation}\label{eqn:Fogler_Kelley}
G \sim \frac{a}{2 L R_0}\exp\left(-\sqrt{\frac{8 T_0}{F a}}\right)\,,
\end{equation}
where $T_0$ is defined by
\begin{equation}\label{eqn:T_0}
T_0 = 1 \,/\, (g a)\,,
\end{equation}
$a$ is the electron localization length, and $R_0$ is specified in Sec.~\ref{sec:Model}. (At such fields mesoscopic conductance fluctuations are small, and so we denote $\langle G \rangle$ simply by $G$.) Our numerical results are consistent with Eq.~\eqref{eqn:Fogler_Kelley}. Note that it can be viewed as the 1D Mott law with the effective temperature~\cite{Shklovskii1973hci, Marianer1992eto} $T_\text{eff} \sim F a$ replacing the ambient temperature $T$. 

Finally, we examine the PDFs of the mesoscopic conductance fluctuations. Such functions can also be studied experimentally, albeit it requires a substantial time and effort.~\cite{Orlov1989spo, Hughes1996dfa} We demonstrate that the PDFs are qualitatively different in the Ohmic and non-Ohmic regimes. Both have asymmetric long tails. However, the Ohmic PDF is skewed towards the low conductances, while the non-Ohmic one towards the high conductances. We explain these differences and show how they evolve as a function of the applied field $F$.

The paper is organized as follows. In Sec.~\ref{sec:Results} we present the summary of our results. In Sec.~\ref{sec:Model} we define the model and introduce our fast algorithm for computing the resistance at a given current.
In Sec.~\ref{sec:DistributionFunctions} we obtain analytical fitting formulas for the PDF of individual hops in the Ohmic and non-Ohmic regimes. We also describe approximate but much faster ``PDF-algorithm'' for computing the net resistances. Section~\ref{sec:G-V_Char} discusses the differences of two averaging procedures: at given current $I$ and at given electric field $F$. Finally, Sec.~\ref{sec:Discussion} contains discussion and comparison with experiments.

\section{Main results}
\label{sec:Results}

In this section we provide a short overview of our principal results for experimentally measurable transport properties.

%
%
\begin{figure}
\includegraphics[width=3.5in]{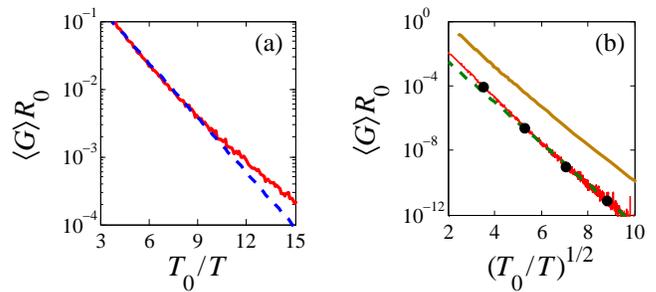}
\caption{Ensemble-averaged Ohmic conductance $\langle G \rangle$ as a function of temperature (the curves with fluctuations): (a) Relatively high $T$. The dashed line is the best fit to the simple exponential law, $\gamma = 1$ and $\Delta = 0.62 T_0$ in Eq.~\eqref{eqn:VRH}. (b) A range of low $T$. The dashed line is a fit to the 1D Mott law, $\gamma = 1 / 2$ and $\Delta = 8.4 T_0$ in Eq.~\eqref{eqn:VRH}. The upper curve is Eq.~\eqref{eqn:RR_G_vs_T}. The dots are the Ohmic limit of the upper four traces in Fig.~\ref{fig:G_nonOhmic}.}
\label{fig:G_Ohmic}
\end{figure}

Figure~\ref{fig:G_Ohmic} shows the dependence of the average Ohmic conductivity $\langle G (0, T) \rangle$ on temperature in an ensemble of
samples of length $L = 250 a$. To test the expected crossover behavior,
we fit the low $T$ data points using Eq.~\eqref{eqn:VRH} with $\gamma = 1/2$, corresponding to the 1D Mott law. We fit higher $T$ using $\gamma = 1$, representing activated transport. In the Mott regime we find $\Delta = 8.4 T_0$. For the activated regime we get $\Delta = 0.62 T_0$. Note the large difference between these values. As far as $\Delta$ is concerned, our numerical results are in a good agreement with the analytical theory of Raikh and Ruzin~\cite{Raikh1989fot} (RR). In the high-$T$ regime it predicts $\Delta = T_0 / 2$. Their low-$T$ formula reads
\begin{equation}\label{eqn:RR_G_vs_T}
G = R_0^{-1} \exp\left(-\sqrt{\nu}\, \frac{T_0}{T}\right)\,,
\end{equation}
where $\nu$ is defined as the solution of the transcendental equation
\begin{equation}\label{eqn:nu_def}
\nu = \frac{2 T}{T_0}\, \ln\left(\sqrt{\nu}\, \frac{L}{a}\right)\,.
\end{equation}
Therefore, RR result for Mott's $\Delta$ is
\begin{equation}\label{eqn:Delta_RR}
\Delta(T) = 2 T_0\, \ln\left(\sqrt{\nu}\, \frac{L}{a}\right)\,.
\end{equation}
Strictly speaking, it is not a constant but a slow function of $T$. In the range of $T$ where the fit to the Mott law was done, it is indeed close to $8.4 T_0$. The large difference between the values of $\Delta$ in the Mott and the activated regime is due to the ``large'' logarithm $\ln (\sqrt{\nu}\, L / a)$.

When the RR formula is plotted alongside our numerical results, it is seen to
exhibit a very similar functional behavior yet a large difference in the absolute value, see Fig.~\ref{fig:G_Ohmic}(b). Despite the fact that we study exactly the same model, see details in Sec.~\ref{sec:Model}, RR's predictions differ from our results by two orders of magnitude. We attribute this discrepancy to the ``subleading'' terms not included in the asymptotic theory of RR.

\begin{figure}
\includegraphics[width=2.6in]{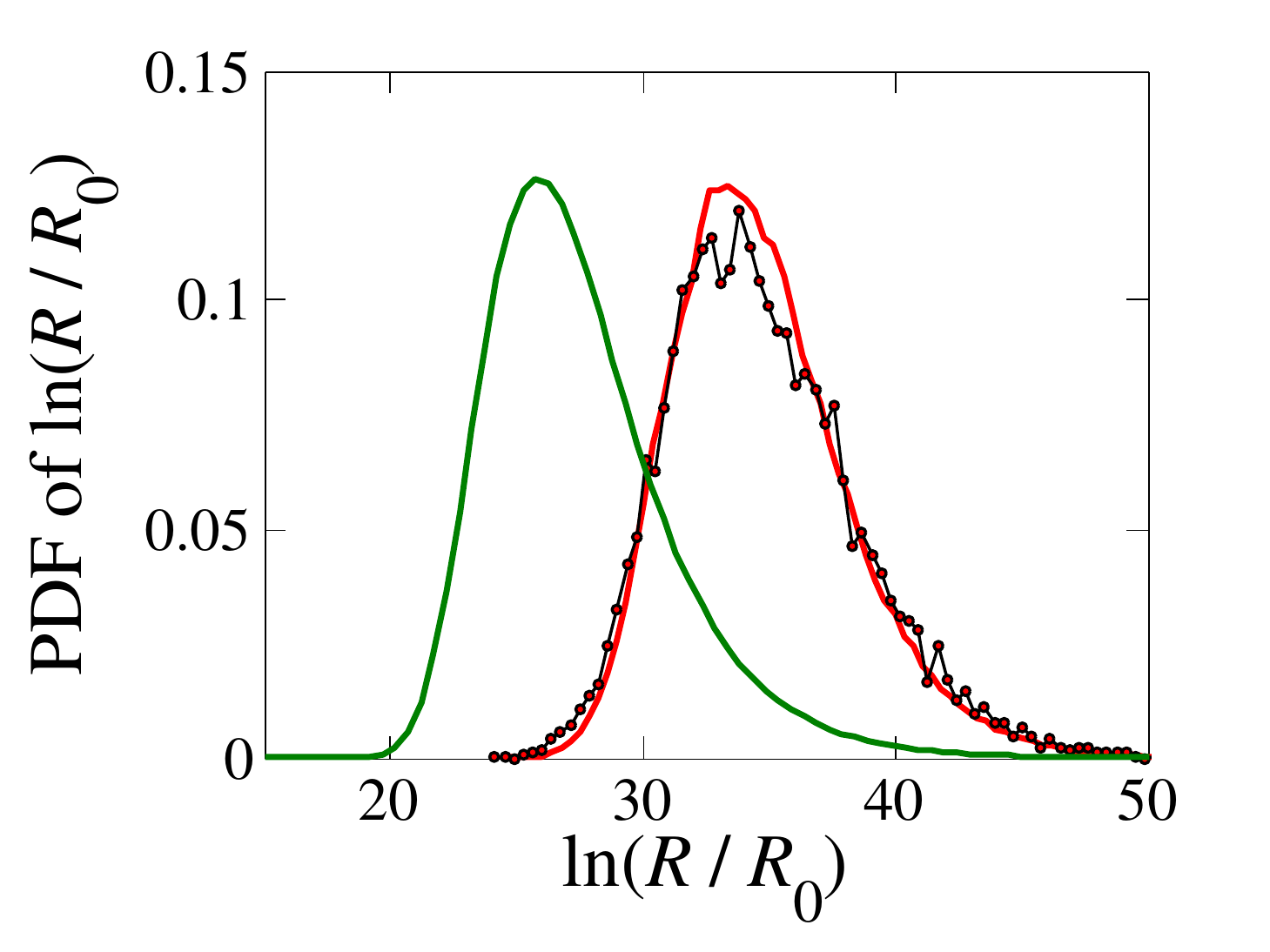}
\caption{The PDF of the logarithm of the total resistance $R$ in the Ohmic limit. The simulations are done for system size $L = 10^3$, localization length $a = 4$, and $u_M = 12.247$. The smooth curve on the right is obtained using the PDF algorithm; the markers correspond to the shortest-path simulation. The leftmost curve is Eq.~\eqref{eqn:RR_PDF_R}.}
\label{fig:R_PDF_Ohmic}
\end{figure}

Next, we present the PDF $P_U(U)$ of the logarithm of the total resistance $U = \ln (R / R_0)$ in the Ohmic limit, Fig.~\ref{fig:R_PDF_Ohmic}, for the same set of wires at temperature $T = T_0 / 75$. Our curves are plotted side-by-side with RR's formula
\begin{gather}
P_U(U) = \sqrt{\nu}\, \exp\left[-\sqrt{\nu}\, \delta U
                           - \exp(-\sqrt{\nu}\, \delta U)\right]\,,
\label{eqn:RR_PDF_R}\\
\delta U \equiv U - (\sqrt{\nu}\, {T_0} / {T})\,.
\label{eqn:Delta_def}
\end{gather}
Again, we see that while the shapes of the curves are practically identical, RR's distribution is centered around a lower value of $U$.
This is consistent with the difference of the $G(T)$ curves described above: ignoring the ``subleading'' terms results in a decreased resistance.

%
%

\begin{figure}
\includegraphics[width=2.9in]{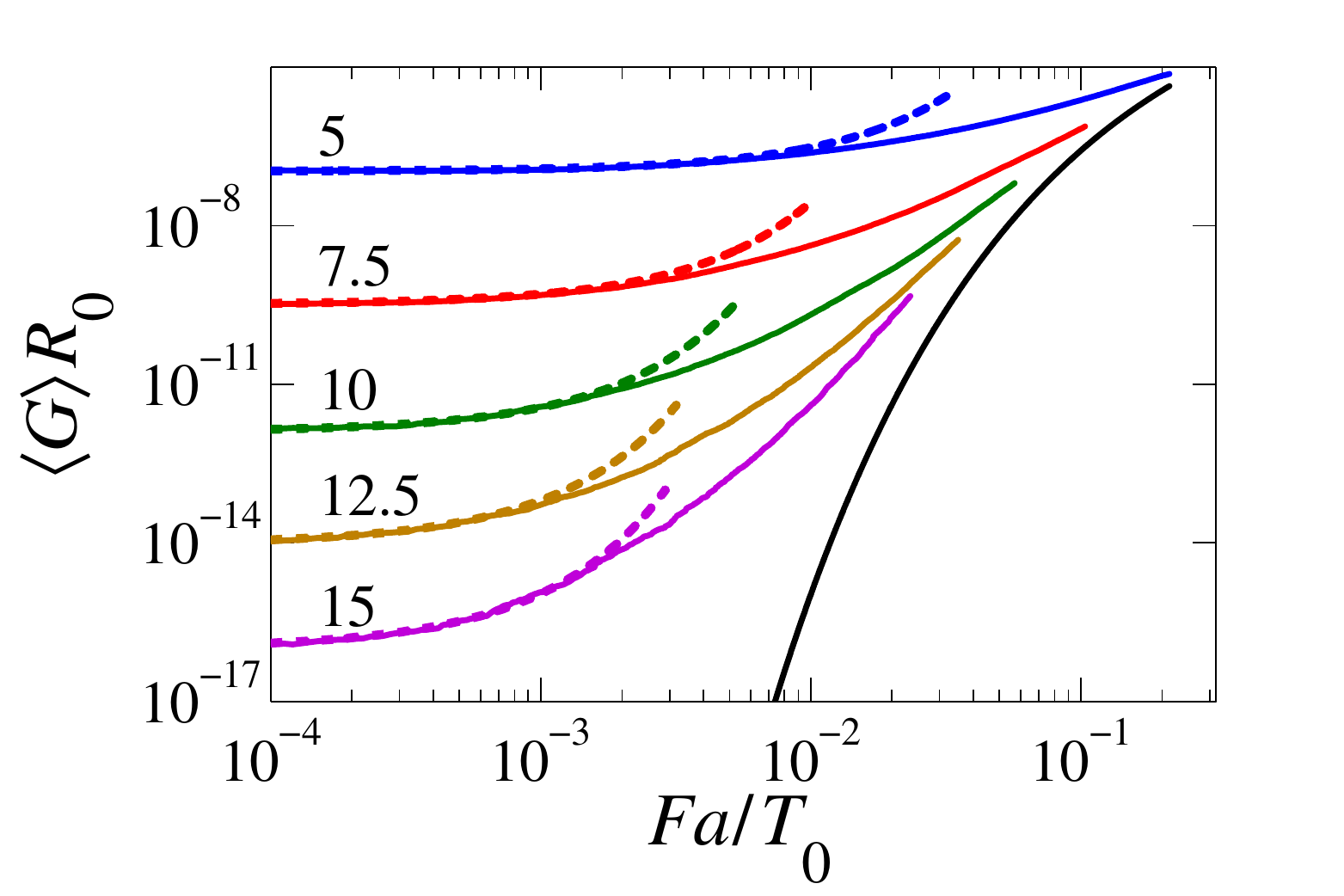}
\caption{Conductance as a function of a scaled electric field $F a / T_0$ (five solid lines on the left). The simulations are done for system size $L = 10^3$ and localization length $a = 4$. The values of $u_M = \sqrt{2 T_0 / T}$ are indicated next to each curve. The fits to Eq.~\eqref{eqn:exponential} used to extract $L_c$ are shown by the dotted lines. The rightmost curve is Eq.~\eqref{eqn:Fogler_Kelley}.}
\label{fig:G_nonOhmic}
\end{figure}

\begin{figure}
\includegraphics[width=2.6in]{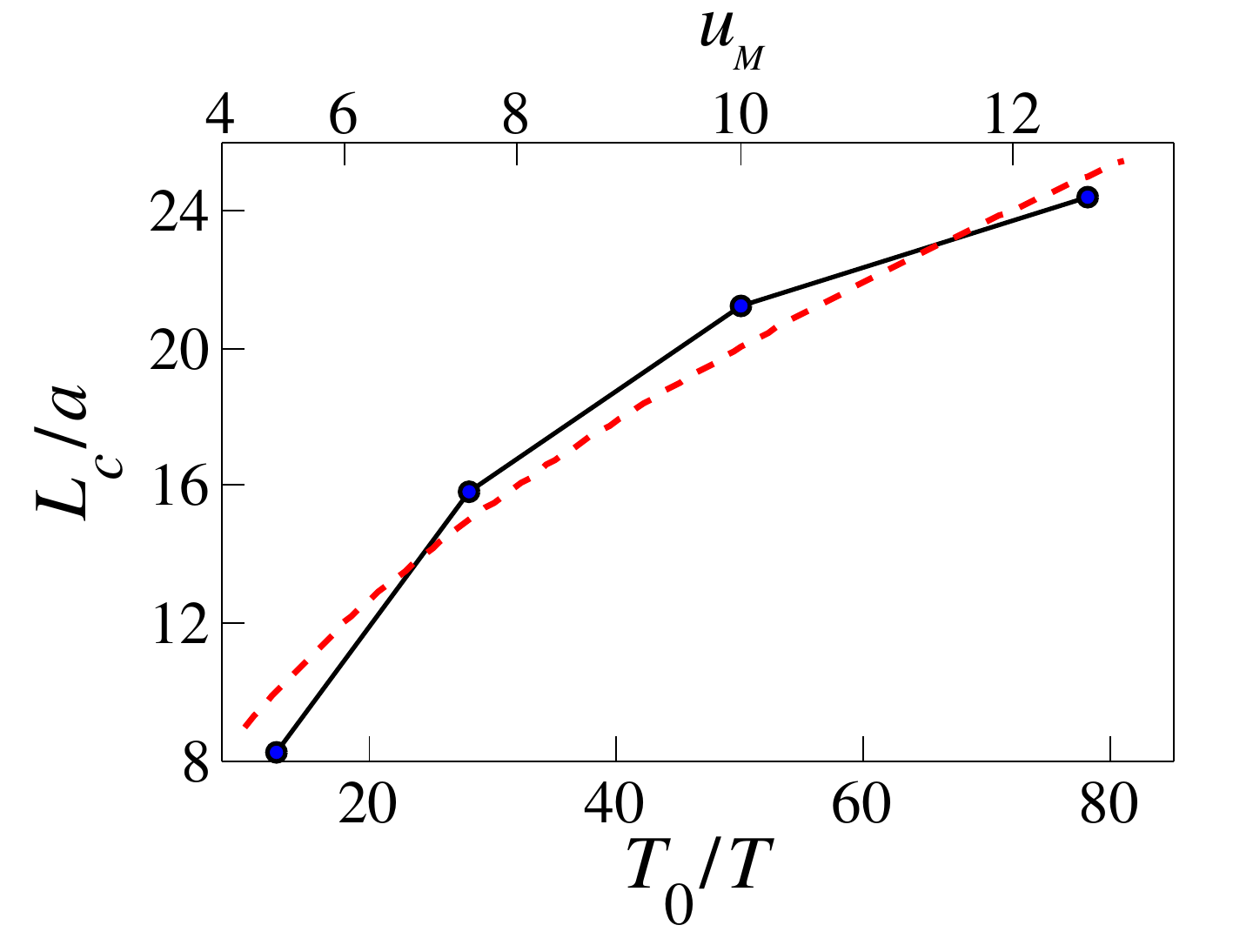}
\caption{Characteristic length $L_c$ [Eq.~\eqref{eqn:exponential}] that determines the non-Ohmic behavior as a function of temperature (dots). For comparison, the dashed curve represents the relation $L_c / a = 1.9 u_M$, which corresponds to a typical hop length.}
\label{fig:Lc}
\end{figure}

\begin{figure}
\includegraphics[width=2.6in]{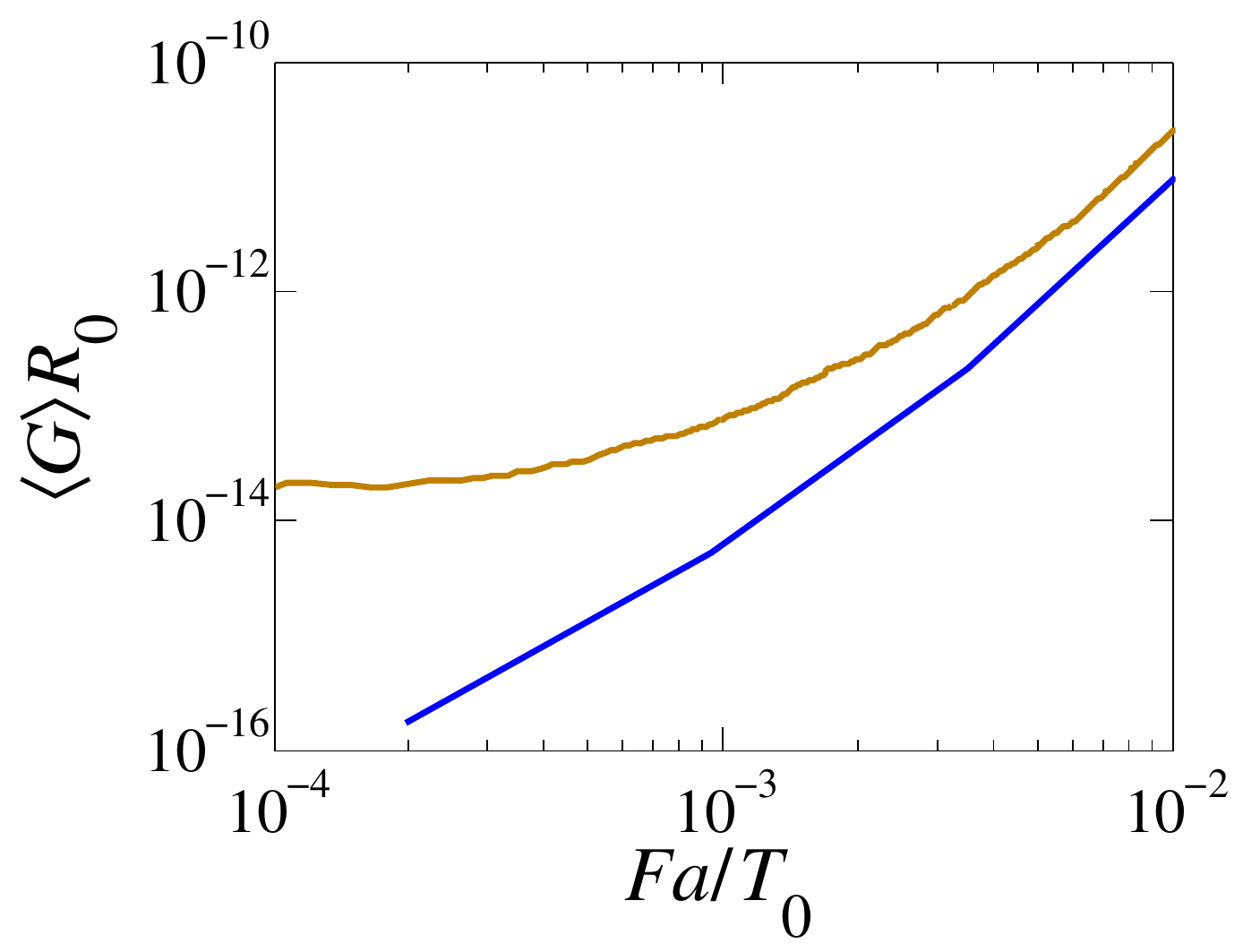}
\caption{Dependence of the conductance on the scaled electric field averaged in two different ways. The upper line is the average conductance, the lower one is the inverse of the average resistance. Simulation parameters are the same as in Fig.~\ref{fig:R_PDF_Ohmic}.}
\label{fig:parallel_series}
\end{figure}

Let us now turn to the non-Ohmic regime. Figure~\ref{fig:G_nonOhmic} illustrates the dependence of the ensemble-averaged conductance as a function of the applied electric field at five different fixed $T$. At low fields the conductance strongly depends on $T$, as the curves originate at points on the vertical axis which differ by many orders of magnitude. [Four of these points are also shown as dots in Fig.~\ref{fig:G_Ohmic}(b).]
All the traces grow monotonically with $F$. Equation~\eqref{eqn:exponential} gives an adequate fit (dotted lines) in a range of low fields. The corresponding $L_c$ are presented in Fig.~\ref{fig:Lc}. We plot them as a function of both the temperature and the ``Mott value", $u_M$, defined as
\begin{equation}\label{eqn:u_M}
u_M\equiv (2 T_0 / T)^{1/2} .
\end{equation}
We see that $L_c \approx 1.9 u_M a$, which is the average hop length. This implies that the average \emph{conductance} is dominated by rare samples that do not contain large breaks, so that the total voltage is distributed roughly equally among all the hops. In contrast, we know that the average \emph{resistance} is determined by typical samples where the breaks are present; the entire voltage is applied to the single most resistive hop, and the size of the non-Ohmic effect is much larger, see Fig.~\ref{fig:parallel_series}. We discuss the difference between average conductance and average resistance in more detail in Sec.~\ref{sec:Discussion}.

%
%

%
%

At large $F$ the rise of the conductance becomes less rapid than exponential and the curves in Fig.~\ref{fig:G_nonOhmic} tend to converge to a common $T$-independent envelope of Eq.~\eqref{eqn:Fogler_Kelley}, confirming the analytical predictions of Fogler and Kelley.~\cite{Fogler2005nov}
At such high electric fields $F$, high-resistance breaks are eliminated not only from rare samples but from typical ones. This can be deduced from the fact that averaging of the conductance $G$ approaches the result of averaging of the resistance $R$ (followed by taking the inverse). As evident from Fig.~\ref{fig:parallel_series}, the two curves indeed approach each other with increasing field. A detailed analysis of this crossover in terms of the PDFs is given in Sec.~\ref{sec:DistributionFunctions}.

%
%

This concludes the summary of our main results. In the next section we define the model and the method of calculation by which they have been obtained.

\section{Model}
\label{sec:Model}
\subsection{VRH resistor network}
\label{sec:Resistor_network}

We model our samples as a network of resistors, as is customary in the VRH theory.~\cite{Miller1960ica} To derive the parameters of this network we proceed as follows. The phonon-assisted transfer of electrons from one localized state (LS) $i$ to another $j$ is characterized by the transition rate 
\begin{equation} \label{eqn:Gamma_ij}
\Gamma_{i \rightarrow j} = \Gamma_0 f_i(1 - f_j) \times \left\{
\begin{array}{ll}
N(\Delta \varepsilon)\,, & \text{if } \Delta \varepsilon > 0\,,
\\
N(|\Delta \varepsilon|) + 1\,, & \text{otherwise}\,,
\end{array}\right.
\end{equation}
where $f_i$ is the occupation factor of $i$\,th LS, $N(\varepsilon)$ is the Bose-Einstein distribution, and $\Delta \varepsilon$ is the energy difference in the hop:
\begin{equation}\label{eqn:Delta_and_Energies}
 \Delta \varepsilon = \varepsilon_j - \varepsilon_i \,,
 \quad \varepsilon_i = \varepsilon_i^0 - e\Phi_i\,.
\end{equation}
Here $\varepsilon_i$ and $\varepsilon_i^0$ are the energy of $i$\,th LS with and without the applied field, respectively, and $\Phi_i$ is its electrostatic potential shift. In a realistic model, the rate prefactor $\Gamma_0$ should have some algebraic dependence on $\Delta\varepsilon$, which counteracts the divergence of $N(|\Delta \varepsilon|)$ at $\Delta\varepsilon \to 0$. However, such $\Delta\varepsilon$ are virtually never important in the VRH transport. For simplicity, we treat $\Gamma_0$ as a constant.

The net current between the LS $i$ and $j$ is given by
\begin{equation}\label{eqn:GammaCurrent}
I_{ij} = -e\,(\Gamma_{i \rightarrow j}-\Gamma_{j \rightarrow i})\,.
\end{equation}
In order to compute $I_{ij}$, one needs to know the occupations factors of all LS. They can be found from the conditions of current conservation
(the so-called Master equation), 
\begin{equation}
\sum_j I_{ij} = 0,
\end{equation}
supplemented by suitable boundary conditions at the source and drain electrodes. Unfortunately, these equations are nonlinear and involve an exponentially large spread of the values of $f_i$. This makes the solution difficult to obtain. It can be done numerically, using some clever iterative techniques.~\cite{Levin1984lth, Yu2001mgf, McInnes1990nco} However, the rate of convergence is slow. We proceed in a different direction, which enables us to map the problem to a resistor network even in the non-Ohmic regime. As a result, we can achieve practically the same speed of simulations in the non-Ohmic regime as in the Ohmic one.

We start by defining the chemical and the electrochemical potentials as follows:
\begin{equation}\label{eqn:Potentials}
\mu_i = T\ln(f_i^{-1}-1) \,, \quad \eta_i = \mu_i - e\Phi_i\,.
\end{equation}
The ``voltage drop" of every $(i, j)$ link is given by the difference of electrochemical potentials $\delta\eta = \eta_i - \eta_j$. In turn, the link resistance is defined by 
\begin{equation}\label{eqn:ResistanceDef}
R_{ij} = {\delta\eta} / {I_{ij}}\,.
\end{equation}
Substituting this into Eq.~\eqref{eqn:GammaCurrent}, one obtains~\cite{Pollak1976apt}
\begin{align}
I &= \frac{2 T}{e R_0} \sinh\left(-\frac{\delta \eta}{2 T} \right)
\exp \left(-\frac{2 x_{i j}}{a}\right)
\notag\\
&\times \exp \left(-\frac{|\varepsilon_i - \eta_i|}{2 T}
- \frac{|\varepsilon_j - \eta_j|}{2 T}
- \frac{|\varepsilon_i - \varepsilon_j|}{2 T}
\right)\,,
\label{eqn:I_pair}
\end{align}
where $x_{ij}$ is the distance between the LS $i$ and $j$, and $R_0 = T / (e^2\Gamma_0)$.

Let us introduce logarithmic variables
\begin{equation}\label{eqn:u_def}
u_{i j} = \ln\, \frac{R_{i j}}{R_0} = u_I + \ln\, \frac{\delta \eta}{T}
\,,
\quad
u_I = \ln\left( \frac{T}{e R_0 I} \right)\,.
\end{equation}
It is easy to see then that if the voltage drop is smaller than $T$ (Ohmic case), the expression for $u_{ij}$ reduces to the well-known 	form~\cite{Shklovskii1984epo}
\begin{equation}\label{eqn:u_Ohmic_def}
u_{ij} = \frac{2 x_{i j}}{a}
+ \frac{|\varepsilon_i - \eta|}{2 T}
+ \frac{|\varepsilon_j - \eta|}{2 T}
+ \frac{|\varepsilon_i - \varepsilon_j|}{2 T}\,.
\end{equation}
Here either $\eta_i$ or $\eta_j$ can be used for $\eta$.

To complete the system of equations, we need a formula for the electrostatic potential $\Phi_i$ [Eq.~\eqref{eqn:Delta_and_Energies}]. It is determined by charges on the source and drain leads, and the perturbation of the electron density inside the wire (given by the occupation factors $f_i$). The relative importance of these contributions depends on the exact geometry of the device. We consider a typical situation where there is a metallic gate positioned parallel to the wire, with $C$ again denoting the capacitance to the gate per unit length of the wire. We further assume that the capacitive coupling to the leads is much smaller and can be neglected. In this case, we find
\begin{equation}
\Phi(x) = -\frac{e n(x)}{C},
\label{eqn:Phi_x}
\end{equation}
where $n(x)$ is the deviation of the local density from equilibrium. Neglecting fluctuations in the local density of states and any correlation effects, we can directly relate $n(x)$ to the local chemical potential, $n(x) = g \mu(x)$, which implies
\begin{equation}
-e \Phi_i = \frac{e^2 g}{C}\, \mu_i =  \eta_i \left(1 - \frac{1}{\epsilon}\right)\,,
\label{eqn:Phi_i}
\end{equation}
where $\epsilon$ is given by Eq.~\eqref{eqn:epsilon}. In comparison, in previous literature it was common to approximate $\Phi_i$ simply by $-F x_i$, i.e., to assume that the electric field in the system is uniform. Although this may be reasonable for a sample of dimension $d > 1$ with bulk leads, it is inappropriate for the specified 1D geometry where the electric field is heavily concentrated at the breaks.

Substituting Eq.~\eqref{eqn:Phi_i} into Eq.~\eqref{eqn:I_pair}, we obtain:
\begin{align}
I &= \frac{2 T}{e R_0} \sinh\left(-\frac{\delta \eta}{2 T} \right)
\exp \left(-\frac{2 x_{i j}}{a}\right)
\notag\\
&\times \exp \left(-\frac{|\varepsilon_i^0 - \eta_i/\epsilon|}{2 T}
- \frac{|\varepsilon_j^0 - \eta_j/\epsilon|}{2 T}\right)
\notag\\
&\times \exp\left(- \frac{|\varepsilon_i^0 - \varepsilon_j^0+(\eta_i-\eta_j)(\epsilon-1)/\epsilon|}{2 T}
\right)\,.
\label{eqn:I_pair_epsilon}
\end{align}
In this equation, all self-consistent field effects are conveniently expressed in terms of the effective dielectric constant $\epsilon$.
Actually, in this paper we focus on the case of weak electron interaction, so that henceforth $\epsilon$ will be replaced by unity. Effect of finite-strength interactions, $\epsilon \neq 1$, will be considered in a separate publication.

To implement the resistor network we proceed as follows. We choose the coordinates of the LS, $0 \leq x_i \leq L$, to be the sites of a chain with unit nearest-neighbor spacing. Their energies $\varepsilon_i^0$ are selected randomly. We draw these energies from the Poisson distribution $P_\varepsilon(z) = g \exp(-g |z|)$ and generate two of them --- one above zero and the other below --- at each internal lattice point. For high currents (small $u_I$) we sometimes generate additional energies at the same lattice point, using the same procedure.

The leftmost lattice point is the source electrode. It has only one LS at the coordinates $(0,0)$, whereas the right end of the sample has many sites at the same $x$-position, equally spaced along the energy axis, see Fig.~\ref{fig:hopping}. This is done in order to simulate the behavior of a metallic drain electrode where there are all energies present.

\subsection{Shortest-path algorithm}
\label{sec:Shortest_path_algorithm}

At this point, we make a crucial approximation, which is, however, conventional in the VRH theory.~\cite{Shklovskii1984epo} We will suppose that there exists a certain path through the network --- the optimal path --- whose conductance is much higher than any other linear path. We can assume then that all the current flows along the optimal path without branching. As we show below, this allows us to devise a fast algorithm for finding such a path and therefore the net resistance of the sample.

%
%
\begin{figure}
  \includegraphics[width=2.6in]{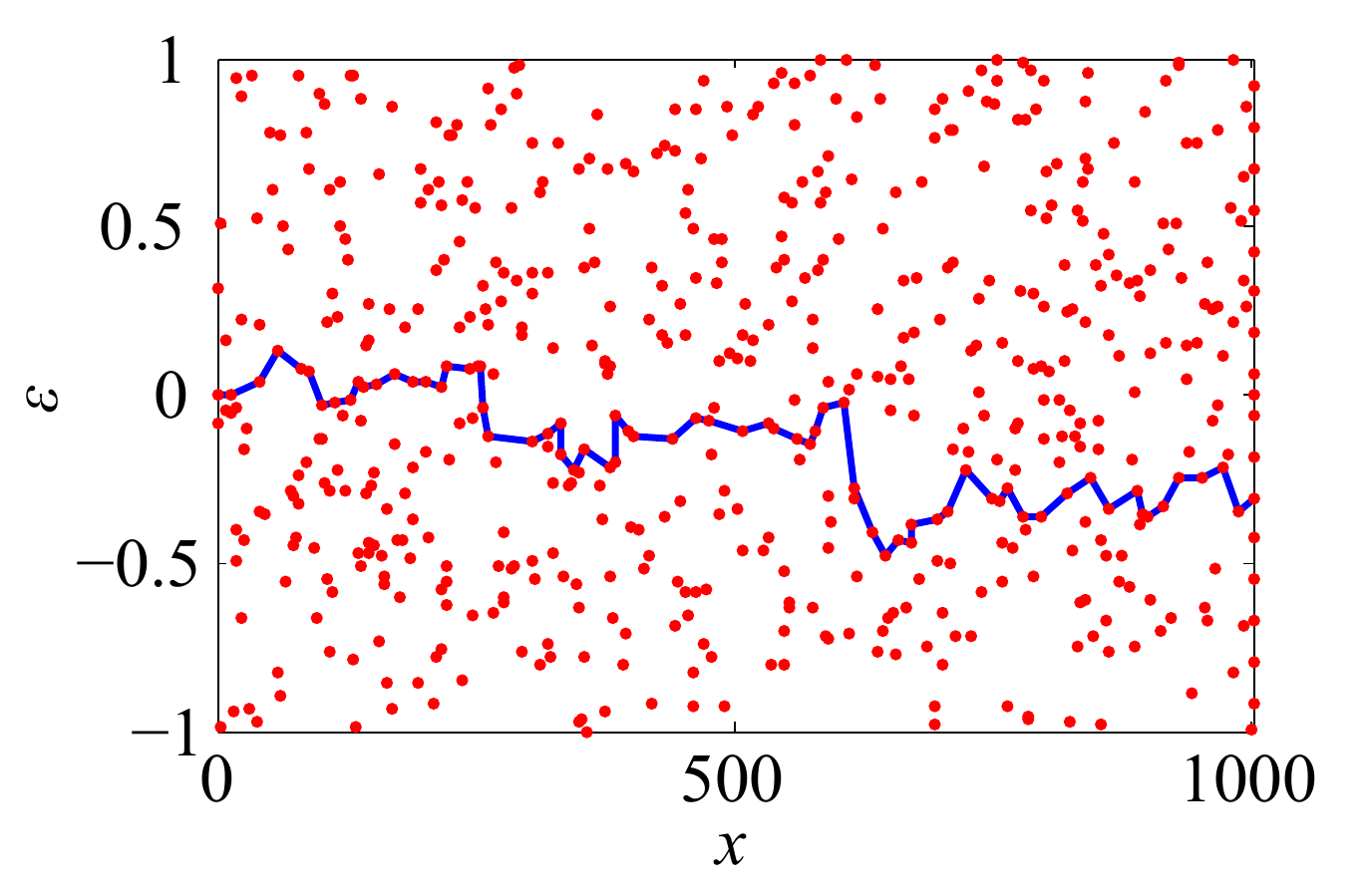}
  \caption{An example of the optimal path in a modestly non-Ohmic regime, $u_I = 25$. The dots represent localized states.}
\label{fig:hopping}
\end{figure}

Using Eq.~\eqref{eqn:I_pair} with $\epsilon = 1$, we can
express the voltage drop $\delta \eta$ in terms of $\eta_i$, the bare site energies $\varepsilon_i^0$ and $\varepsilon_j^0$,
and $u_I$. To this end we define axillary variables $t$ and $q$:
\begin{align}
t &= (\varepsilon_j^0 - \eta_i) / T\,,
\label{eqn:t}\\
q &= \frac{2 x_{i j}}{a} 
   + \frac{|\varepsilon_i^0 - \varepsilon_j^0|}{2 T} 
   + \frac{|\varepsilon_i^0 - \eta_i|}{2 T}
   + \frac{t}{2} - u_I\,.
\label{eqn:q}
\end{align}
Only $q < 0$ are physically allowed, which means that there is a certain maximum
current that can flow through the given link. If so, the voltage drop in question
is
\begin{numcases}{\frac{\delta\eta}{T} = }
 -\ln\left(1 - e^{q}\right)\,, & if $e^{q} > 1 - e^t$,
\label{eqn:u_nonOhmic_I}\\
  \ln\left(1 + e^{q - t} \right)\,, & otherwise.
\label{eqn:u_nonOhmic_II}
\end{numcases}
One can show that this cumbersome expression is reduced to the familiar Eq.~\eqref{eqn:u_Ohmic_def} in the low-current limit, $u_I \to \infty$. Indeed, in this case, Eq.~\eqref{eqn:u_nonOhmic_I} applies for $t < 0$, while
Eq.~\eqref{eqn:u_nonOhmic_II} for $t \geq 0$. Note that Eq.~\eqref{eqn:u_Ohmic_def} is independent of $u_I$, as is appropriate in the Ohmic regime.

We can use the above equations to find the optimal path through the sample. This is the path which would require the lowest voltage (difference in the electrochemical potential between the ends of the sample) for a given current. To do so we use the well-known Dijkstra algorithm~\cite{Dijkstra1959ano} to calculate the minimum ``cost" of getting from the source to the drain. Here the cost is the total voltage $V$. Similarly, the cost $c_i$ of getting to site $i$ on the optimal path is
\begin{equation}
c_i = -\eta_i\,.
\label{eqn:cost}
\end{equation}
The algorithms starts by assigning zero cost to the source $(0, 0)$ and infinite cost to all other sites. Thereafter the spanning tree of the lowest-cost sites is grown iteratively. Initially, the tree consists of only the source site. At each iteration, a site of the lowest cost among those that are still outside the tree is added to the tree. The costs of sites $j$ outside the tree are relaxed (updated) according to the rule
\begin{equation}
c_j^{(n + 1)} = \min \big(c_i^{(n)} + \delta\eta,
                c_j^{(n)}\, \big)\,.
\label{eqn:cost_update}
\end{equation}
Here $c_i^{(n)}$ is the cost of site $i$ at $n$\,th iteration. The cost increment $\delta\eta$ in Eq.~\eqref{eqn:cost_update} is computed using Eqs.~\eqref{eqn:u_nonOhmic_I} and \eqref{eqn:u_nonOhmic_II}. The process terminates when any of the LS located on the drain electrode are reached.
In Fig.~\ref{fig:hopping}, one can see an example of an optimal path found by our algorithm in a modestly non-Ohmic regime.

In the Ohmic VRH problem, the Dijkstra algorithm has been used in Ref.~\onlinecite{He2003spa}. In that regime each link has a fixed cost. Here we are using the Dijkstra algorithm in an unconventional situation where the cost $\delta\eta = \delta\eta(c_i)$ of a given link is not a constant but a nonlinear function of the cost of the earlier sites in the tree. A potentially troublesome point is that in the course of iterations we retain only the lowest cost so far. We effectively assume that for any $i$ and $j$
\begin{equation}\label{eqn:sum_of_costs_assumption}
\min c_j = \min \big(c_i + \delta\eta(c_i)\big) = \min c_i
                                        + \delta\eta(\min c_i)\,.
\end{equation}
Let us show that this equation is satisfied, which implies that our algorithm works correctly at arbitrary current. First of all, by our earlier assumption the current does not branch, and so the current through any link of the optimal path must be exactly $I$. Second, a sufficient condition for validity of Eq.~\eqref{eqn:sum_of_costs_assumption} is $\partial c_j / \partial c_i \geq 0$. That is, increasing $c_i$ by taking a less optimal path to the $i$\,th site would not help to decrease $c_j$. In view of Eq.~\eqref{eqn:cost}, the last condition can be written as
\begin{equation}\label{eqn:cost_requirement}
\frac{\partial}{\partial\eta_i}\, \delta\eta \leq 1\,.
\end{equation}
We need to examine the two possible cases represented by Eqs.~\eqref{eqn:u_nonOhmic_I} and \eqref{eqn:u_nonOhmic_II}. In the former, we get
\begin{equation}\label{eqn:Dijkstra_I}
\frac{\partial}{\partial\eta_i}\, \delta\eta  = -\frac{e^q}{2(1 - e^q)} [\sgn(\varepsilon_i - \eta_i) + 1] \leq 0 < 1\,.
\end{equation}
In the latter, we obtain
\begin{equation}\label{eqn:Dijkstra_II}
\frac{\partial}{\partial\eta_i}\, \delta\eta  = \frac{e^{q - t}}{2(1 + e^{q - t})}[1 - \sgn(\varepsilon_i - \eta_i)] < 1\,.
\end{equation}
In both cases inequality~\eqref{eqn:cost_requirement} is satisfied, which means that our algorithm does find the optimal path.

In the course of simulations, the resistance of every link on this path as well as their total sum are saved for further analysis. Repeating the process over many disorder realizations, we obtain the PDFs and the averages of desired transport properties, discussed in more detail below.

\section{Distribution functions}
\label{sec:DistributionFunctions}

In this section we review analytical predictions regarding the functional form of the PDF of link resistances and compare them with the simulation results. In both Ohmic and non-Ohmic cases we are able to make the two to agree by introducing a few refinements in the analytical formulas and by adjusting numerical coefficients therein.

\subsection{Ohmic case}
\label{sec:Ohmic_case}

%
%
\begin{figure}
\includegraphics[width=3.0in]{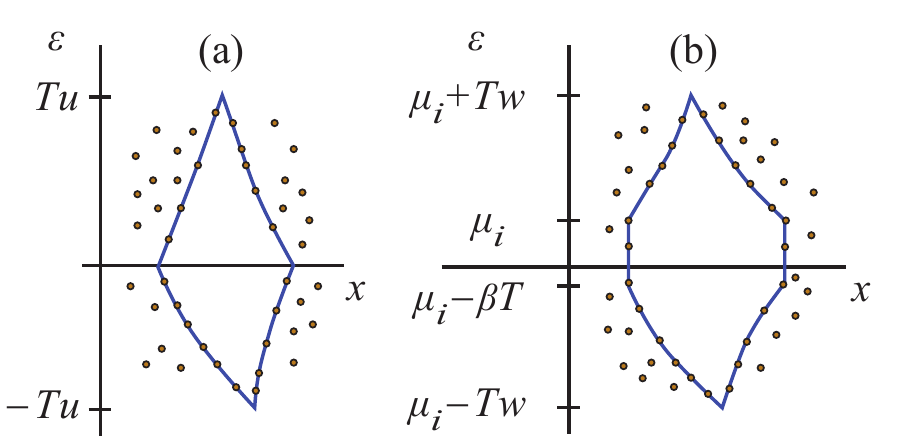}
\caption{Geometry of (a) Ohmic (b) Non-Ohmic break in the energy-position space. The dots represent localized states.}
\label{fig:break}
\end{figure}

We start by discussing the Ohmic case: $u_I\rightarrow\infty$.
According to previous theoretical studies, notably Refs.~\onlinecite{Lee1984vrh}, \onlinecite{Serota1986nao}, and \onlinecite{Raikh1989fot}, the logarithm of the average resistance of a link is on the order of the Mott value $u_M$. Links with $u \gg u_M$ are exponentially rare; however, they act as bottlenecks and the total resistance depends on them. In order for such high-resistance links to exist, the optimal path has to encounter regions in the energy-position ($x$-$\varepsilon$) space that are empty of LS. Using the method of optimal fluctuation, RR~\cite{Raikh1989fot} showed that the leading asymptotic behavior of the PDF of the breaks has the form
\begin{equation}\label{eqn:RR_PDF}
P(u) = -\frac{d}{du}\exp[-g A(u)]\,,
\end{equation}
where $A(u)$ is the smallest possible area of a break with given $u$ in the $x$-$\varepsilon$ space. Equation~\eqref{eqn:RR_PDF} is due to the Poisson distribution of the LS in the $x$-$\varepsilon$ space. The shape that
attains the minimal area depends on whether the break is Ohmic ($u < u_I$) or non-Ohmic ($u - u_I \gg 1$). For the former case, RR showed that the break is diamond-shaped with the width $u a / 2$ and the height $2 u T$, see
Fig.~\ref{fig:break}(a). This entails the quadratic dependence
\begin{equation}\label{eqn:gA}
g A(u) = (u / u_M)^2\,.
\end{equation}
Later, taking into account shape fluctuations of the break along its perimeter, Ruzin~\cite{Ruzin1991fso} proposed a refined formula
\begin{equation}\label{eqn:Ruzin_PDF}
P(u) = C_0 \exp(2 B u / u_M) \times g A'(u)\exp[-g A(u)].
\end{equation}
While $C_0$ is determined essentially by the normalization of $P$, analytical calculation of the coefficient $B$ is challenging. Ruzin gave a rough estimate $B \approx \sqrt{2}/3 \approx 0.5$. In this study, we calculate $B$ numerically. Indeed, from the example of the optimal path shown in Fig.~\ref{fig:hopping}, it is clear that the voids around the long hops hardly ever look like ``diamonds" (or ``hexagons", see below). This means that even though the RR theory provides the basis for understanding the behavior of $P(u)$, numerical simulations are critical in order to calculate it accurately.

At each $L$ the functional form of the $P(u)$ is expected to depend only on the dimensional ratio $u / u_M$. By running simulations at different combinations of $a$, $g$, and $T$, we convinced ourselves that this is indeed correct, for the exception of very small $u$ where lattice discreteness starts to matter. Fortunately, such $u$ are irrelevant for the macroscopic transport properties as they do not determine the resistance. Thereafter we fixed $a = 4$, $g = 1 / 3$, and $T = 0.01$, which yields the characteristic temperature $T_0 = 3 / 4$ and the Mott parameter $u_M = 12.247$, cf.~Eq.~\eqref{eqn:u_M}. To ensure we are in the Ohmic regime $u_I = 200 \gg u_M$ was used.

\begin{figure}
  \includegraphics[width=3.3in]{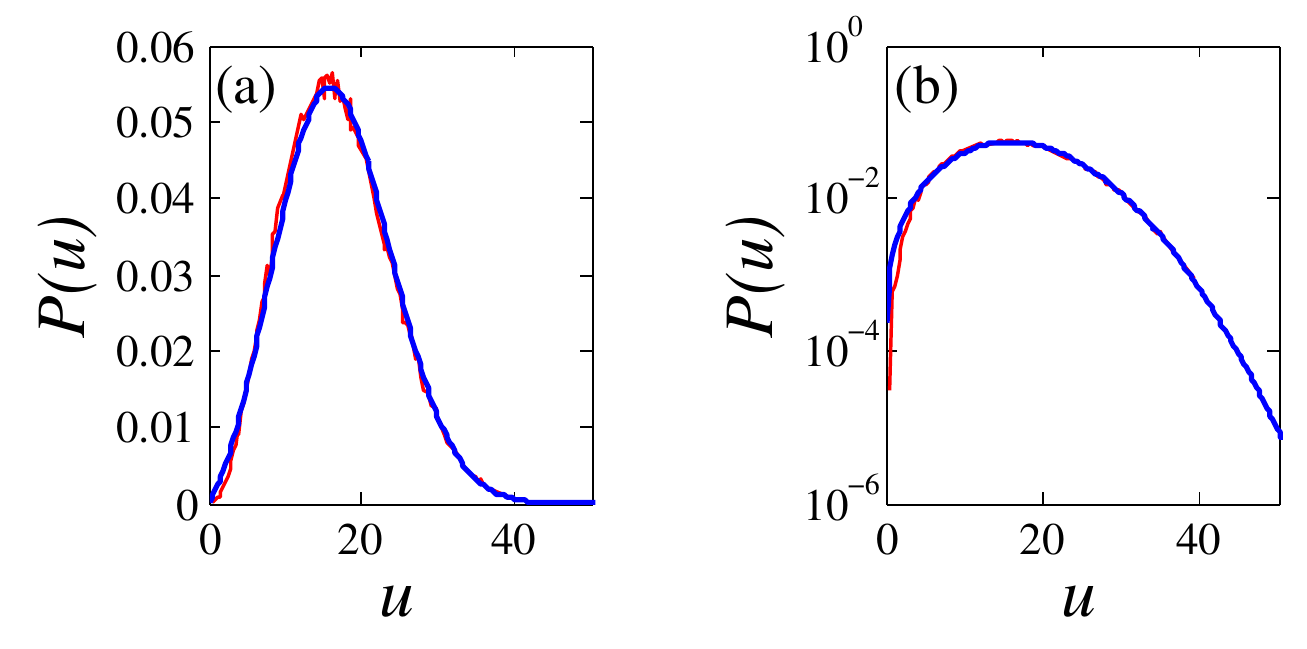}
  \caption{Numerical results for $P(u)$ in the Ohmic regime shown on (a) linear and (b) logarithmic scale. The simulation parameters are the same as in Fig.~\ref{fig:R_PDF_Ohmic}, e.g., $u_M = 12.247$  (thin line). The small fluctuations are of statistical origin. Equation~\eqref{eqn:Ruzin_PDF} with $B = 0.9$ is represented by the smooth thick line.}
\label{fig:Ohmic_PDF}
\end{figure}

For each $L$ in the set $L = 100, 200, 400, 500,$ and $1000$ we generated many realizations of 1D wires, respectively, $20000$, $10000$, $5000$, $4000$, and $2000$. Anticipating the finite-size effects, these numbers were chosen in order to have the same total number $2 \times 10^6$ of LS at each $L$. We found optimal paths through the samples and created the PDFs of the link resistances. We fitted such PDFs to Eq.~\eqref{eqn:Ruzin_PDF} using $B$ as a single adjustable parameter. The quality of the fits was rather good, see an example in Fig.~\ref{fig:Ohmic_PDF}. Furthermore, even though Eq.~\eqref{eqn:Ruzin_PDF} is meant to apply at $u \gg u_M$, it fits our numerical results for $u \lesssim u_M$ as well.

Interestingly, we found that $B$ slowly but systematically increases with $L$. When plotted as a function of $1 / L$, it was seen to vary linearly, tending to a constant for large $L$. We believe that the reason for this finite-size effect is the following: due to the source electrode being at zero energy, the resistance of the first link is typically lower than average. In shorter samples, where the total number of hops through the sample $N_u$ is about ten or so [see Eq.~\eqref{eqn:N_u} below], it impacts the PDF. As the samples get longer, $N_u$ increases and this first hop does not influence the overall PDF any more. To get the value of coefficient $B$ in the thermodynamic limit, we used linear extrapolation to $L = \infty$. Our final estimate is
\begin{equation}\label{eqn:B}
B = 0.92 \pm 0.02\,,
\end{equation}
approximately twice larger than that of Ref.~\onlinecite{Ruzin1991fso}.

%
%

Two characteristic measures of the width of the distribution are its mode and its average. For $P(u)$ they are given by, respectively,
\begin{align}
u_{\max} &= \frac12 \left( B + \sqrt{B^2 + 2}\, \right)\,u_M
 = (1.30 \pm 0.02)\, u_M\,,
\label{eqn:u_max}\\
\langle u \rangle &= \int\limits_0^{\infty} u P(u) d u
                  = (1.39 \pm 0.02)\, u_M\,.
\label{eqn:u_bar}
\end{align}
As expected, both are the order of the Mott parameter $u_M$. One more important quantity is the average number $N_u$ of links on the path.
It determines the relation between $P(u)$ and the probability density of breaks per unit length of the wire $\rho(u)$:
\begin{equation}
\rho(u) = \frac{N_u}{L}\, P(u)\,. 
\label{eqn:rho}
\end{equation}
Since the width of each link is not smaller than $(a / 2) u$,
cf.~Eq.~\eqref{eqn:I_pair}, $N_u$ can be estimated from below as $(2 L / a) / \langle u \rangle \approx 1.4 L / au_M$. According to our
simulations, the actual $N_u$ is approximately twice larger:
\begin{equation}\label{eqn:N_u}
N_u = (3.04 \pm 0.07)\, \frac{L}{u_M a}\,.
\end{equation}
Besides RR~\cite{Raikh1989fot} and Ruzin~\cite{Ruzin1991fso}, the calculation of $P(u)$ was previously attempted by Ladieu and Bouchaud.~\cite{Ladieu1993csiI} They reported $u_{\max}$ and $ \langle u \rangle$ that differ from our Eqs.~\eqref{eqn:u_max} and \eqref{eqn:u_bar} by $30$-$40$\% In fact, we were unable to
verify that statement because the main equation of Ref.~\onlinecite{Ladieu1993csiI} has no solution. As written, that equation  does not conserve probabilty. Consequently, we believe that our results constitute the first reliable calculation of function $P(u)$.

\subsection{Non-Ohmic case}
\label{sec:Non-Ohmic_case}

Let us now discuss the breaks in the non-Ohmic regime. Unlike the diamonds of the Ohmic case, the non-Ohmic breaks are hexagonal, see Ref.~\onlinecite{Fogler2005nov} and Fig.~\ref{fig:break}, with area
\begin{gather}
g A(u) = \frac{w^2 + w \beta}{u_M^2}\,,
\quad w = u_I + \ln(1 - e^{-\beta})\,,
\label{eqn:A_nonOhmic_I}\\
\beta = \frac{\delta \eta}{T} = e^{u - u_I}.
\label{eqn:beta}
\end{gather}

The width of the break in the real space is $w a / 2$. Note that at $u < u_I$ we
have $w a / 2 \simeq u a / 2$, which is the width of the Ohmic break. The
combination $\beta T$, which is equal to the electrochemical potential drop
across the break, gives the the height of the middle part of the break in the
$x$-$\varepsilon$ space.

In order to account for the possible perimeter corrections to $P(u)$, we consider the following trial form:
\begin{align}
P(u) &= C_0 \exp\left[2 B \frac{w}{u_M} + C \left(\frac{\beta}{u_M}\right)^D\right]
\notag\\
&\times g A^\prime(u) \exp\left[-g A(u)\right]\,.
\label{eqn:P_nonOhmic}
\end{align}
Here the contribution of the top and bottom parts of the perimeter is modeled after Eq.~\eqref{eqn:Ruzin_PDF}. It is proportional to the length of such parts $\sim w$ and the coefficient $B$. The contribution of the side walls of the break, of length $\beta T$, is written differently. Indeed, Ruzin's argument~\cite{Ruzin1991fso} suggests that they give no contribution at all. In fact, we found it necessary to include a correction albeit with a smaller exponent $D = 0.5$. We have no other justification for this exponent except that it provides a good fit to the numerical $P(u)$, see below. The explicit formula for $P(u)$ can be derived from Eqs.~\eqref{eqn:A_nonOhmic_I}--\eqref{eqn:P_nonOhmic} by the straightforward differentiation with respect to $u$. However, it is cumbersome and we do not write it here. Equation~\eqref{eqn:P_nonOhmic} applies for $u - u_I \gg 1$ and $u_I \gg u_M$. It refines the corresponding expression for $P(u)$ in Ref.~\onlinecite{Fogler2005nov} where the first (subleading) exponential term was not included. The Ohmic and non-Ohmic formulas, Eqs.~\eqref{eqn:Ruzin_PDF} and \eqref{eqn:P_nonOhmic}, match at $u - u_I \sim 1$.

\begin{figure}
\includegraphics[width=3.3in]{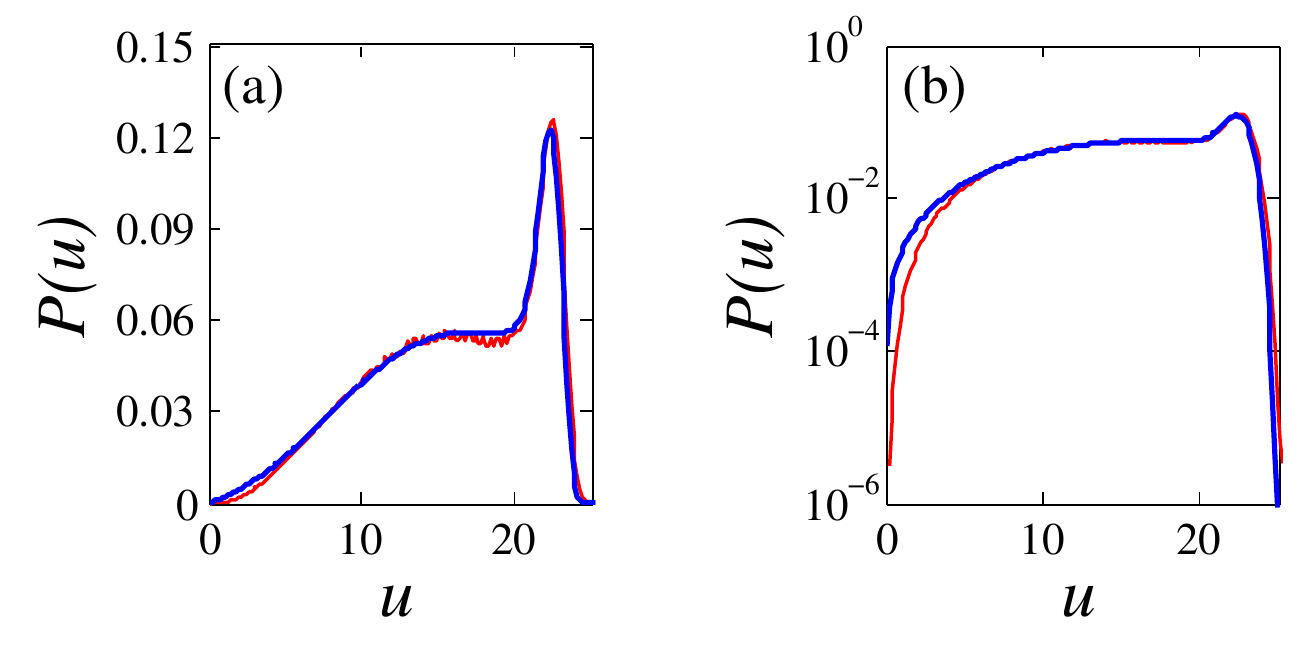}
\caption{Numerical results for $P(u)$ for finite current, $u_I = 20$, shown on (a) linear and (b) logarithmic scale (thin line). The small fluctuations are of statistical origin. The fitting formula~\eqref{eqn:P_nonOhmic} with $u_M = 12.247$, $B = 0.9$, $C = 0.75$, and $D = 0.5$ is represented by the thick line.}
\label{fig:nonOhmic_PDF}
\end{figure}

Equation~\eqref{eqn:P_nonOhmic} predicts that $P(u)$ decays as a Gaussian at $u_M < u < u_I$ and as an exponential of the exponential at $u > u_I$. In between, it exhibits a narrow peak of width $\delta u \sim \ln (u_M^2 / u_I)$ near the non-Ohmic threshold $u = u_I$. For parameters chosen in Fig.~\ref{fig:nonOhmic_PDF} this peak is so pronounced that it already dwarfs the ``Ohmic'' maximum at $u = u_{\max}$. The reason for its appearance is similar to that discussed in a three-dimensional case.~\cite{Shklovskii1976nhc} This narrow peak is due so-called ``soft'' links that used to have resistances $u \gtrsim u_I$ in the Ohmic regime. Such links are similar to forward-biased diodes: their conductance increases exponentially with the electrochemical potential drop $\delta \eta$. When a finite current is made to flow across the wire, such links self-generate $\delta \eta$ large enough to push their resistance back to an immediate vicinity of the non-Ohmic threshold $u \approx u_I$.

The soft links are realized when the energies at their endpoints satisfy a certain inequality, which can be derived from Eq.~\eqref{eqn:I_pair} or looked up in Table I of Ref.~\onlinecite{Pollak1976apt}. Therefore, not all links are soft. There also also ``hard'' links, which are similar to reverse-biased diodes, whose resistance does not change much with $\delta \eta$. These links are never included in the optimal path because they are simply not able to support the necessary current $I$. The peculiar shape of $P(u)$ that follows from these arguments is nicely confirmed by simulations, which we now briefly describe.

The simulation procedure in the non-Ohmic regime is practically identical to the Ohmic case with one exception: we have to put more than two energy sites at each lattice point $x$. The reason for this is that for high currents (and, therefore, high voltages), as the electron moves through the sample, it hops onto LS with lower energies, see Fig.~\ref{fig:hopping}. The addition of extra LS is done to ensure that there are LS for the electron to hop onto, otherwise the path would not be found. We also have to increase the range of the energies on the electrode for exactly the same reason. The simulation was conducted at $u_I = 35, 30, 25,$ and $20$. Two values of $u_M$ are used: $12.247$ (same as above) and $20$ (obtained by adjusting the temperature but keeping $g = 1 / 3$ the same). The fit of the numerical $P(u)$ to Eq.~\eqref{eqn:P_nonOhmic} for $u_M = 12.247$ can be seen in Fig.~\ref{fig:nonOhmic_PDF} and it is quite good at all but very small $u$ (which are irrelevant, see the note above).

%
%

\subsection{Distribution of the net resistance}
\label{sec:PDF-algorithm}

\begin{figure}
\includegraphics[width=3.2in]{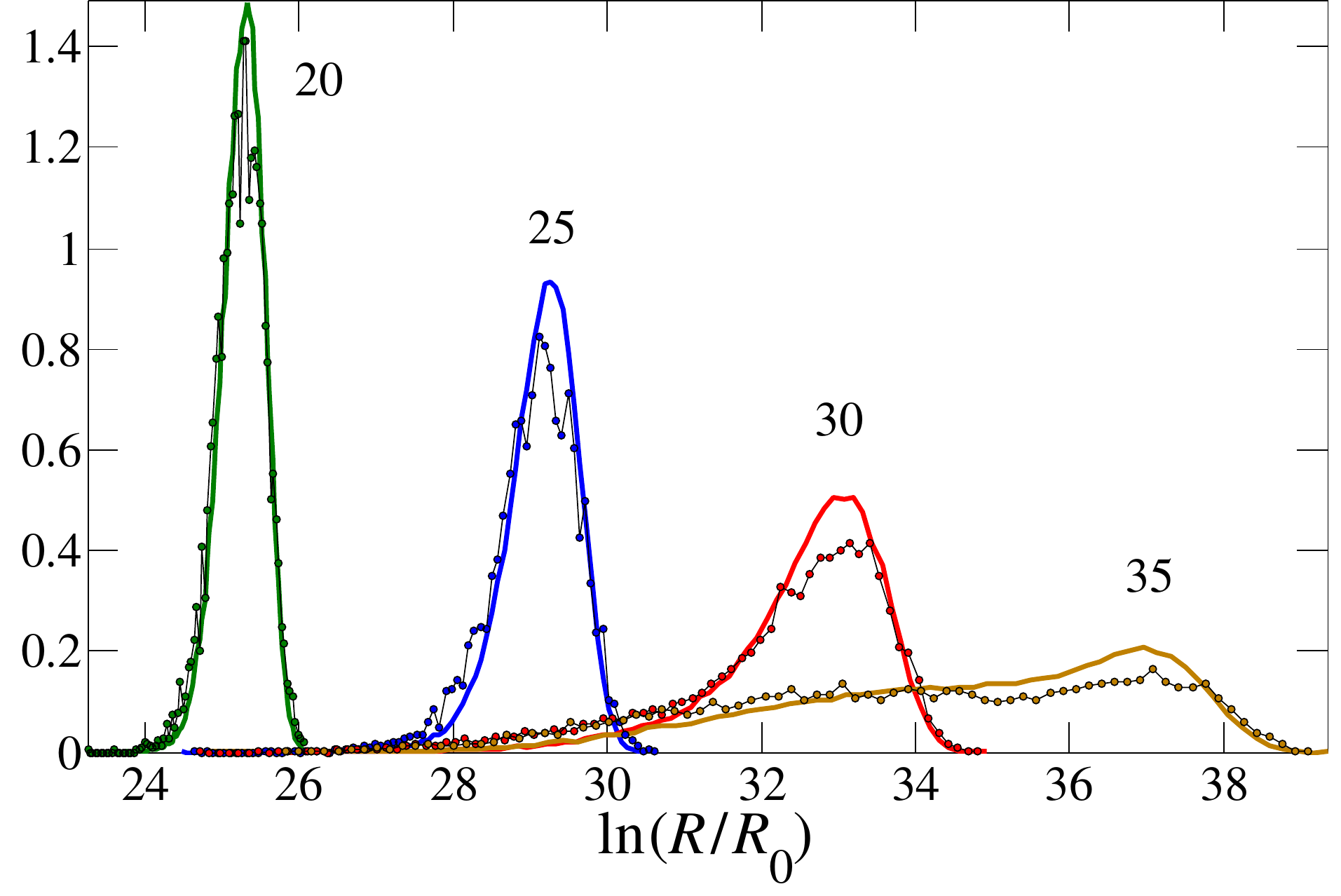}
\caption{The PDF of the logarithm of the total resistance $R$. The values of $u_I$ are indicated next to each curve. The simulation parameters are the same as in Fig.~\ref{fig:R_PDF_Ohmic}. The smooth curves are obtained using the PDF algorithm, the markers are from the shortest-path simulations.}
\label{fig:R}
\end{figure}

Besides studying the distribution of individual hops, we also investigated the statistics of the net resistance $R$. In Fig.~\ref{fig:R} we present the a sequence of four PDF's of $U \equiv \ln(R / R_0)$ obtained from our shortest-path simulations. From one curve to the next the current increases by the same factor of $\exp(5)$. A qualitative difference from the PDF for the Ohmic case (Fig.~\ref{fig:R_PDF_Ohmic}) is immediately apparent.
The Ohmic PDF is skewed to the right, towards the large resistances.
In contrast, the non-Ohmic curves skewed the opposite way. This difference is due to the response of $P(u)$ (the PDF of individual links) to the rise in current. In both Ohmic and non-Ohmic regimes the net resistance of the system is determined by the largest breaks. But in the non-Ohmic case there is almost a hard cutoff $\approx u_I$ on the largest possible $u$ (Fig.~\ref{fig:nonOhmic_PDF}). In other words, breaks with $u \gtrsim u_I$ are effectively eliminated,~\cite{Fogler2005nov} making the large-resistance side of the PDFs of $\ln(R / R_0)\gg u_I$ drop sharply as well.

Another result of removing the highly resistant links is the PDF's approach to the Gaussian shape. By reducing the spread of the link resistances, it brings the system closer to the conditions at which the central-limit theorem is obeyed. This can be seen in Fig.~\ref{fig:R}, where the curves become narrower and more Gaussian at lower $u_I$.

Also plotted in Fig.~\ref{fig:R} are PDFs obtained by an approximate but much faster method, which utilizes our analytical formulas for $P(u)$.
We call this the PDF-algorithm. The idea is as follows.~\cite{Raikh1989fot}
The resistance of the system is given by the sum over all links,
\begin{equation}
R = R_0 \sum_{i = 1}^{N_u} e^{u_i}\,.
\label{eqn:R_sum}
\end{equation}
Under the assumption that the link resistances are independent random variables, each with the same PDF $P(u)$, it can be shown that
\begin{align}
P_U(U) &= \frac{1}{2\pi}\int \exp \left(U - i t e^U \right)
 \mathcal{G}(t) d t\,,
\label{eqn:P_U}\\
\mathcal{G}(t) &= \exp \bigg\{ L\int \rho(u)[\exp(i t e^u) - 1] d u \bigg\}\,,
\label{eqn:G_t}
\end{align}
This is equivalent to the formulas given by RR in Refs.~\onlinecite{Raikh1987tfi} and \onlinecite{Raikh1989fot}. For convenience of the reader, we include a quick derivation. For independent variables the cumulants of the sum are equal to the sum of the cumulants.~\cite{vanKampen2008} To calculate the latter we notice that the number of breaks of size $(u, u + d u)$ has the average value $d N(u) = L \rho(u) d u$. The actual number is random and has the Poisson distribution. Therefore, its contribution to $n$\,th cumulant of $R / R_0$ is $e^{n u} d N$.
The total cumulant is
\begin{equation}
\kappa_n = L \int \rho(u) e^{n u} d u\,.
\label{eqn:cumulant}
\end{equation}
Reconstructing the characteristic function $\mathcal{G}(t)$ from the cumulants in a standard way,~\cite{vanKampen2008} we obtain Eq.~\eqref{eqn:G_t}. Taking its Fourier transform and making the change of variable from $R$ to $U$, we recover Eq.~\eqref{eqn:P_U}.

Certainly, the resistances of the links are not truly uncorrelated; however, since $R$ is dominated by the largest breaks, which are rare and well-separated, this should be a good approximation. Note that in Ref.~\onlinecite{Ladieu1993csiI} an attempt was made to include
correlations between adjacent links. As mentioned above, it does not compare well with our simulations.

In practice, even a numerical integration of the strongly oscillating functions in Eqs.~\eqref{eqn:P_U} and \eqref{eqn:G_t} is difficult. We found it easier to directly implement Eq.~\eqref{eqn:R_sum} instead. To this end we draw $u_i$ from the distribution $P(u)$ using a Monte-Carlo sampling (the usual acceptance-rejection algorithm). After $N_u$ [Eq.~\eqref{eqn:N_u}] of such resistances are generated, the total resistance of the wire is obtained by summing them. Figure~\ref{fig:R} illustrates that the PDFs obtained from the shortest-path simulations and from the PDF-algorithm are in a good agreement. The curves produced by the latter are much more smooth because we could apply it to a larger number of disorder realizations: $10^5$.

%
%

\section{Conductance-voltage characteristics}
\label{sec:G-V_Char}

Having studied the statistics of individual hops that contribute to the 1D transport, we can now move to the analysis of macroscopic transport properties. In experiment, such transport properties are measured either as a function of current or as a function of voltage. In the former case, the ensemble averaging gives the average resistance $\langle R \rangle$; in the latter --- the average conductance $\langle G \rangle$. If a large number of nominally identical wires is available simultaneously, this can be done in a single measurement, connecting them, respectively, in series and in parallel.~\cite{Khavin1998slo} Otherwise, one can try to create the members of an ensemble one by one by varying gate voltage or other parameters of a single wire.~\cite{Hughes1996dfa}

Since our shortest-path algorithm is formulated at a constant current (i.e., constant $u_I$), one may naively think that it is able to provide only the distribution of resistances. This is not so. Let us show that the PDFs of conductances and resistances are uniquely related even in the non-Ohmic regime.

%
%
\begin{figure}
  \includegraphics[width=2.6in]{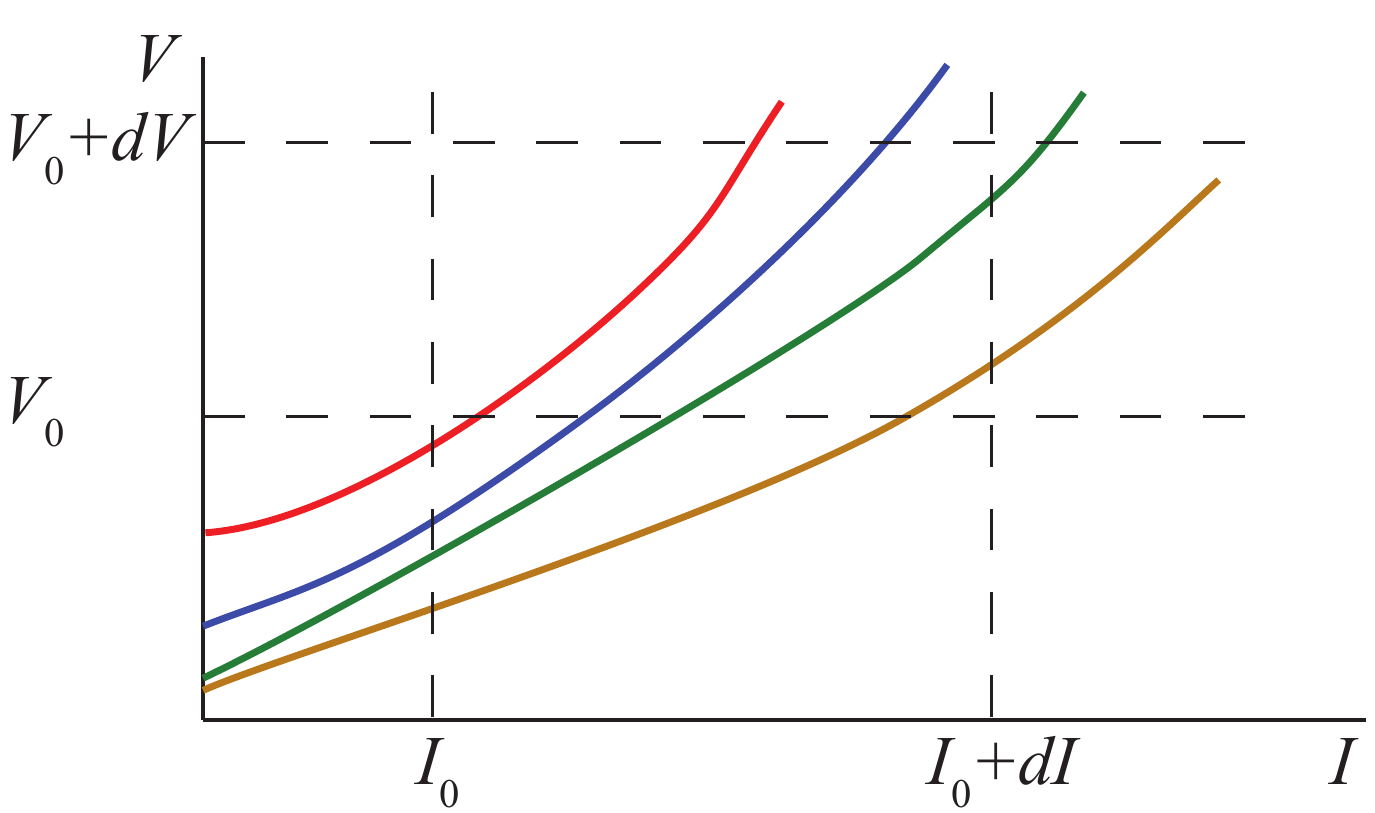}
  \caption{A sketch of $V$--$I$ curves for an array of different wires. The continuity equation \eqref{eqn:flow_conservation} follows from the conservation of the number of curves piercing the differential area element bounded by the dashed lines.}
\label{fig:V_vs_I}
\end{figure}

Denote the PDF of having a given total voltage $V$ at a fixed current $I$ by $P_V(V | I)$ and the PDF of having a given current $I$ at a fixed total $V$ by $P_I(I | V)$. By inspecting the $V$-$I$ curves sketched in Fig.~\ref{fig:V_vs_I}, we can write down the following continuity equation:
\begin{align}
\frac{\partial}{\partial I}\, P_V(V | I)
 + \frac{\partial}{\partial V}\, P_I(I | V) = 0\,.
\label{eqn:flow_conservation}
\end{align}
Integrating with respect to voltage, we get
\begin{equation}\label{eqn:Int_V}
P_I(I | V)  = - \frac{\partial}{\partial I}\int_0^{V}P_V(V'|I)dV'\,.
\end{equation}
As an application, let us show how the average conductance $G_V$ at a given fixed voltage $V$,
\begin{equation}\label{eqn:def_G}
G_V = \int_0^\infty P_I(I|V) \frac{d I}{R}\,,
\quad R = \frac{V}{I}\,,
\end{equation}
can be calculated.

In view of Eq.~\eqref{eqn:Int_V}, $G_V$ can also be written as
\begin{equation}\label{eqn:def_G1}
G_V = -\int_0^\infty\frac{I  d I}{V} \frac{\partial}{\partial I}\int_0^V P_V(V'|I) d V'\,.
\end{equation}
We integrate this by parts and change the notation for the measure in the second integral from $P_V(V'|I) d V'$ to $P_R(R'|u_I) d R'$. We arrive at the formula
\begin{equation}\label{eqn:G_V}
\frac{G_V}{R_0^{-1}} =
\frac{T}{V} \!\! \int_{-\infty}^\infty \!\! \frac{d u_I}{e^{u_I}}
\!\! \int_0^\infty \! \Theta \!\left(\frac{V e^{u_I}}{T} - \frac{R'}{R_0} \right)
P_R (R' | u_I) d R'
\end{equation}
for the desired average conductance at a fixed voltage. It is easy to see that in the Ohmic limit, $V \to 0$, Eq.~\eqref{eqn:G_V} coincides with the average conductance at a fixed current, $\int P_R(R' | \infty) d R' / R'$, as expected.

To evaluate $G_V$ as a function of $V$ one needs to know $P_R(R'|u_I)$. We obtained it by the following procedure. We divided
the interval of $V$ we are interested in into a number of bins.
We took an interval of $u_I$ from $5$ to about $u_I = 3u_M$ and in turn divided it into equidistant steps $u_I(j)$, $1 \leq j \leq N_I = 1000$, spaced by $\Delta u_I$. For each $u_I(j)$ we generated $N_\text{sam} = 200$ samples, i.e., sets of $N_u$ individual $u$'s, drawn from the distribution $P(u)$ using the acceptance-rejection algorithm. We converted the integrals in Eq.~\eqref{eqn:G_V} into discrete sums,
\begin{equation}\label{eqn:G_sim}
\frac{G_V}{R_0^{-1}} =
\frac{ \Delta u_I}{V N_\text{sam}} \sum_{j = 1}^{N_I}
\frac{T}{e^{u_I(j)}}
\sum_{i = 1}^{N_\text{sam}} \!\Theta\! \left(\frac{Ve^{u_I(j)}}{T}
- \frac{R_i(j)}{R_0} \right),
\end{equation}
where $R_i(j)$ is the total resistance of $i$\,{th} set for a given $j$, and then evaluated them numerically.

The simulations were done for $u_M = 5$, $7.5$, $10$, $12.5$, and $15$. The control parameter was $T$ while all other values --- $a$, $g$, $L$, and $N_u$ --- remained the same. Later we realized that in the non-Ohmic regime the number of hops $N_u$ gradually increased with current. Equation~\eqref{eqn:N_u} remains accurate only for $u_I > u_M$. Therefore, only $u_I > u_M$ points were included when plotting the five curves in Fig.~\ref{fig:G_nonOhmic}.

Alternatively, $G_V$ can be reduced to a numerical quadrature, which this time contains no oscillating integrands. This is possible because
$G_V$ is dominated by large conductances, for which the saddle-point approximation in Eq.~\eqref{eqn:P_U} is legitimate. After a straightforward derivation, one obtains
\begin{align}
\frac{G_V}{R_0^{-1}} &=
\frac{T}{V} \int_{-\infty}^\infty \frac{d u_I}{e^{u_I}}
\int_0^\infty \Theta \!\left(\frac{V e^{u_I}}{T} - J_1 \right)
\notag\\
&\times  \sqrt{\frac{J_2}{2 \pi}}\, \exp \big(J_1 t + J_0) d t\,,
\label{G_V_quad}\\
J_n &= N_u \int_0^\infty P(u) \left[
\exp(n u - t e^u) - \delta_{n, 0}
\right] d u\,,
\label{J_n}
\end{align}
where $n = 0, 1, 2$, and $\delta_{i j}$ is the Kronecker symbol. All these integrals are rapidly converging, so that their numerical evaluation should cause no difficulty. However, we deemed the quality of the curves shown in Fig.~\ref{fig:G_nonOhmic} sufficient [these curves were obtained from Eq.~\eqref{eqn:G_sim}]. Therefore, we did not pursue this alternative method.

\section{Discussion}
\label{sec:Discussion}

At this point, let us recapitulate our findings. To the best of our knowledge we presented the first reliable calculation of the statistics of resistances in 1D VRH network, both in Ohmic and non-Ohmic regimes. Comparing with the previous theoretical work, we showed the importance of the correction to the PDF $P(u)$ proposed in Ref.~\onlinecite{Ruzin1991fso}. We demonstrated that without this ``subleading'' term the conductance could be significantly overestimated, see Fig.~\ref{fig:G_Ohmic}. Figure~\ref{fig:hopping} further illustrates the importance of such corrections by showing that there are no obvious diamond-like or hexagonal voids in the energy-position space invoked in the derivations of the leading asymptotic behavior.~\cite{Raikh1989fot, Fogler2005nov} 

Next, our calculations have verified the earlier analytical predictions~\cite{Fogler2005nov} that large breaks are progressively eliminated at higher voltage, and that the PDF of resistances becomes more narrow, see Fig.~\ref{fig:R}. This disappearance of highly resistive hops equalizes different samples, making the averages of parallel and series setups of the wires approach the same value.

Let us now turn to experiments. Unfortunately, we could not find a clear evidence of the predicted behavior in published literature. A dedicated experiment to probe mesoscopic conductance fluctuations in non-Ohmic regime is desired as it was not on the agenda in previous studies of 1D VRH. At least two other caveats must also be kept in mind. First, most of ``1D'' electron systems studied experimentally were not truly one-dimensional. They either consisted of many parallel chains~\cite{Aleshin2005cbt, Aleshin2004hci} or had multiple subbands~\cite{Khavin1998slo, Gershenson2000hee, Gao2006cao} or were bulk samples with a large aspect ratio.~\cite{Orlov1989spo, Hughes1996dfa} Such systems may behave as effectively 1D but only at low enough $T$. Finally, our model of disorder where LS are treated as points in the energy-position space may or may not be relevant for some of these experiments (see more below).

Turning to some specific examples, we consider first the measurements done on  polydiacetylene single crystals,~\cite{Aleshin2004hci} which are quasi-1D materials. The Ohmic transport is consistent with 1D VRH behavior, showing a crossover from a simple exponential at relatively high temperatures,
$\ln G \approx -\Delta_h / 2 T$, to a stretched exponential $\ln G \approx -(\Delta_l / T)^\gamma$ with $\gamma = 0.5$--$0.75$ at low $T$. Just as in our simulations, there is a substantial difference between $\Delta_h$ and $\Delta_l$. For instance, in sample S1 $\Delta_h = 320\,\text{K}$ and 
$\Delta_l = 2570\,\text{K} \approx 8 \Delta_h$. In the same sample
at high electric fields Eq.~\eqref{eqn:Fogler_Kelley} is observed, with
$8 T_0 / a  = 0.049\,\text{eV}/\text{nm}$ (in our notations).
Assuming that $T_0 \approx \Delta_h$, this gives a reasonable estimate of the
localization length $a = 4.3\,\text{nm}$. At modest fields, the transport data were fitted to Eq.~\eqref{eqn:exponential} and $L_c$ was extracted. It was seen to have the same temperature dependence $L_c \propto T^{-0.5}$, as in our simulations. Moreover, the numerical value of $L_c$ is close to what we find. For example, $L_c = 32.5\,\text{nm}$ at $T = 25\,\text{K}$ in the experiments,~\cite{Aleshin2004hci} which can be compared to $L_c \sim 1.9 a \sqrt{2 T_0 / T} = 40\,\text{nm}$ that we find, cf.~Fig.~\ref{fig:Lc}. 

Next, let us consider another experiment, which was done on arrays of GaAs quantum wires.~\cite{Khavin1998slo} The dependence of $G$ on $F$ and $T$ that we have calculated here is in a reasonable agreement with some of those experimental results but some strong deviations are also apparent. For example, in the simulations the range of activated behavior in the Ohmic regime spans at best two decades in $G$. In the experiment, it is much wider (three decades), and occupies most of the temperature range $T > 0.2\,\text{K}$ where $G$ was reported. We were able to fit the experimental $G(0, T)$ only by imposing a rather strong power-law dependence of the prefactor: $G_0 \propto T^{2.5}$. From such a fit we obtained $T_0 = 6.2\,\text{K}$ (in our notations).

\begin{figure}
\includegraphics[width=2.9in]{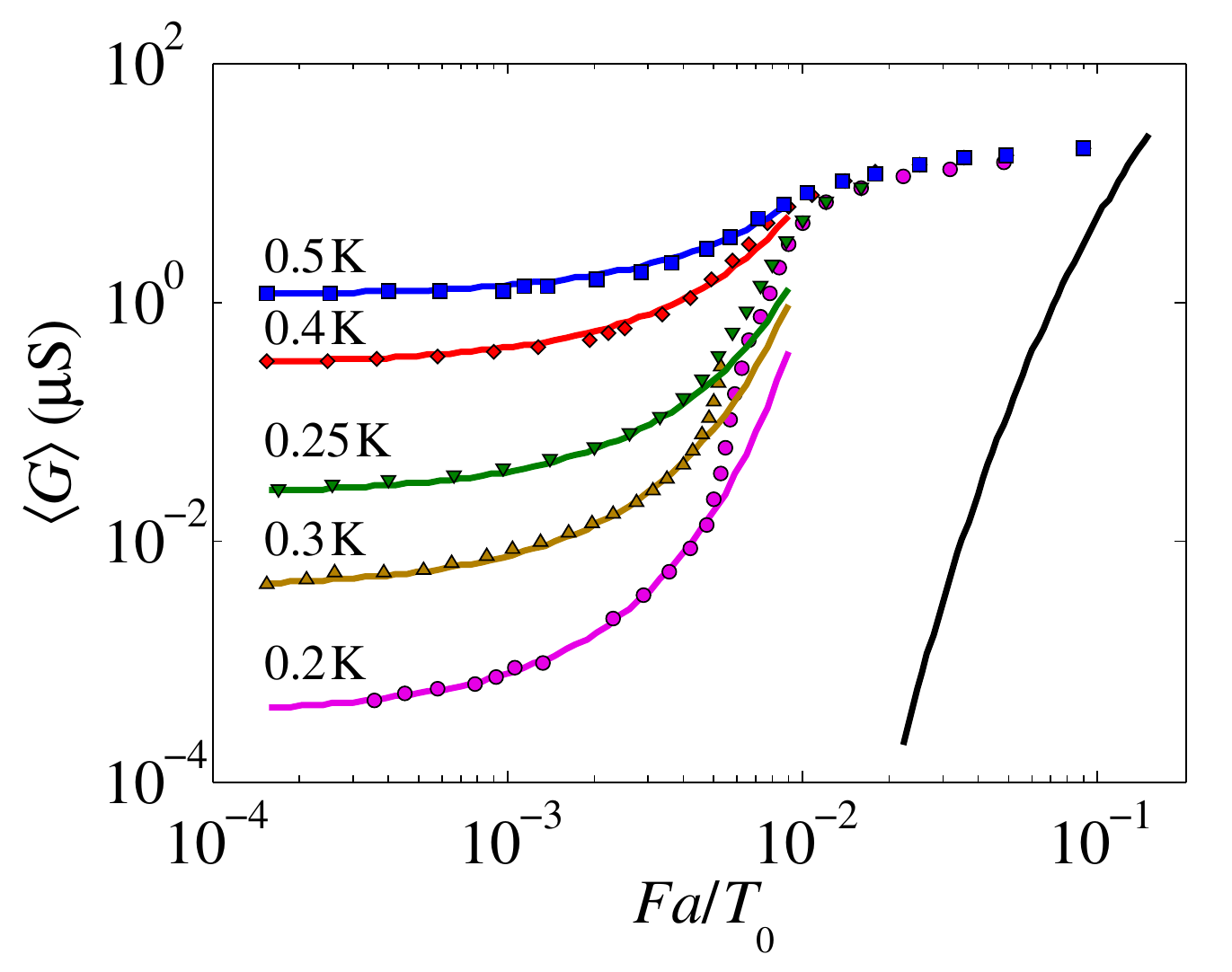}
\caption{Conductance of sample 1 of Ref.~\onlinecite{Khavin1998slo} as a function of the scaled electric field $F a / T_0$  (markers). Here $T_0 = 6.2\,\text{K}$ is determined from the best fit of Eq.~\eqref{eqn:RR_G_vs_T} to the Ohmic conductance (not shown) and $a = 0.4\,\mu\text{m}$. Temperature (in K) is indicated next to each data set. The best fits to Eq.~\eqref{eqn:exponential} are shown by the lines. The rightmost curve is Eq.~\eqref{eqn:Fogler_Kelley}. The prefactor $G_0$ is chosen such that the relation between Eq.~\eqref{eqn:exponential} and the uppermost data trace (corresponding to $u_M \approx 5$) is similar to that in Fig.~\ref{fig:G_nonOhmic}.}
\label{fig:G_nonOhmic_Khavin}
\end{figure}

In the non-Ohmic regime, the initial rise of $G$ with $F$ is again exponential over approximately one decade, see Fig.~\ref{fig:G_nonOhmic_Khavin}. However the behavior of parameter $L_c$ in this exponential law was deemed to be surprising in Ref.~\onlinecite{Khavin1998slo}. Therefore, let us discuss it. Physically, $L_c$ is the distance between ``critical hops" in a sample, i.e., those highly resistive links that generate the dominant portion of the total voltage. In a typical sample, length $L_c$ has to be much larger than the average hop length $u_M a$. In fact, at low $T$ one would naively expect $L_c$ to be of the order of the sample length $L$. This is because in a typical sample all the voltage drops on a single break. At higher $T$, where the activated transport is observed, the voltage is shared by many breaks,~\cite{Raikh1989fot} and so $L_c$ is supposed to decrease exponentially. However, this is not what was observed. At low $T$, two out of three samples measured in Ref.~\onlinecite{Khavin1998slo} had $L_c \approx  L / 50$, while $L_c$ of the third was about $L / 10$. As $T$ was increasing, $L_c$ was decreasing but rather slowly, perhaps, as $T^{-1 / 2}$.

%
%

In light of our findings, this behavior of $L_c$ is \emph{not} surprising. The above reasoning does not take into account that the measurements were done not on a single wire but on several hundreds of them, connected in parallel. It is logical to assume that some wires conducted much better than others because they happened to have no breaks. These wires could short out the wires which were poor conductors, reducing the net $L_c$ down to the typical hopping length.

We now demonstrate explicitly that $L_c$ extracted from our model has numerical values and functional behavior similar to what was measured experimentally. Our $L_c$, which was found by fitting the low-voltage part of $G(F)$ curves in Fig.~\ref{fig:G_nonOhmic} to Eq.~\eqref{eqn:exponential} is plotted in Fig.~\ref{fig:Lc}. The intervals of $T_0/T$ are different in our simulation and the experiment; however, there is a small overlap. For our leftmost point, $T_0 / T = 12.5$ we have $L / L_c \approx 30$, similar to the numbers quoted above.

The problem arises when we consider the high-field behavior reported in Ref.~\onlinecite{Khavin1998slo}. Experimental $G(F, T)$ curves tend to approach a common $T$-independent limit, as in our calculations, Fig.~\ref{fig:G_nonOhmic}. However, this limit is strongly underestimated by our Eq.~\eqref{eqn:Fogler_Kelley}, see Fig.~\ref{fig:G_nonOhmic_Khavin}. 
While we do not know the origin of this discrepancy, it is possible that the different behavior seen in the two experiments is just another example of a dilemma, which has a long history in the VRH literature. Previously, it was discussed mostly in the context of bulk materials where majority of experiments have been done so far. However, it is tempting to make a comparison with our 1D case because the VRH exponent of the Efros-Shklovskii law in any dimension nominally coincides with the 1D Mott law exponent $\gamma = 1 / 2$.

The essence of the dilemma is as follows. There are a number of systems where non-Ohmic behavior does follow Eqs.~\eqref{eqn:exponential} and \eqref{eqn:Fogler_Kelley} that we have observed in our simulations. However, this is usually the case when parameter $\Delta$ in Eq.~\eqref{eqn:VRH} is large, say, tens or hundreds of K. Very different and still poorly understood behavior occurs when $T_0$ is relatively small (according to one study,~\cite{Zhang1998noe} when $\sqrt{T_0 / T} \lesssim 12$). The high-field nonlinearities in this second group are much stronger. In the extreme cases, the $I$--$V$ characteristic was determined to be $S$-shaped,~\cite{Ladieu1996dti, Stefanyi1997anc} which led to hysteretic conductivity jumps by orders of magnitude~\cite{Ladieu1996dti, Ovadia2009epd} and circuit oscillations.~\cite{Stefanyi1997anc, Note_on_NDC}
Interestingly, in systems that show conductivity jumps the Ohmic conductance shows a simple activation rather than VRH behavior.~\cite{Ladieu1996dti, Ovadia2009epd} 

It has become common~\cite{Wang1990eat, VanderHeijden1992ncv, Stefanyi1997anc, Zhang1998noe, Gershenson2000hee, Marnieros2000dpn, Leturcq2003hhe, Galeazzi2007hee, Ovadia2009epd, Altshuler2009jic} to attribute strong nonlinearity and $S$-shaped $I$--$V$ to electron overheating. It is assumed that $G$ is the function of the electron temperature $T_e$, which can be much higher than the ambient temperature $T$. A phenomenological equation is postulated,
\begin{equation}
\dot{Q} = G F^2 = \alpha (T_e^\beta - T^\beta)\,,
\label{eqn:T_e}
\end{equation}
where $\alpha$ and $\beta$ are adjustable constants. (Usually, $4 < \beta < 8$.) This equation is supposed to represent the balance between the Joule heat delivered into electron system from the external field and the heat transferred from electrons to phonons. Surprisingly, this equation has been shown to provide an accurate description of some VRH systems, including the the one we are trying to make comparison to.~\cite{Khavin1998slo, Gershenson2000hee}

By itself, the idea of hot electrons is not objectionable. Actually, our Eq.~\eqref{eqn:Fogler_Kelley} can be viewed as the 1D Mott law with the electron temperature $T_e \sim F a$ (similar to Refs.~\onlinecite{Shklovskii1973hci} and \onlinecite{Arkhipov1994fde}). The difficulty is that the required $T_e$ is unusually large. Indeed, let us define the length $L_{\text{e-ph}} = T_e / F$. It has the physical meaning of a characteristic distance over which an electron must be accelerated by the external field to gain the extra energy $T_e \gg T$. In our model, where LS are treated as points, the largest achievable $L_{\text{e-ph}}$ is of the order of $a$. Electrons cannot propagate farther without suffering an exponential decay. Yet to get a stronger $I$--$V$ nonlinearity than predicted by our Eq.~\eqref{eqn:Fogler_Kelley}, $L_{\text{e-ph}}$ must exceed $a$. For example, to reproduce the high-field part of the data shown in Fig.~\ref{fig:G_nonOhmic_Khavin}, we need perhaps $L_{\text{e-ph}} \sim 10 a$.

In principle, $L_{\text{e-ph}} \gg a$ is possible if the disordered system is a granular metal or equivalently, an array of random-sized quantum dots. In this case the upper bound on $L_{\text{e-ph}}$ is presumably set by the size of metallic grains, while the exponential decay length $a$ is much smaller, being suppressed by weak tunneling between the grains. The granular-metal model can also explain a wide range of the activated Ohmic behavior as a manifestation of the Coulomb blockade. Finally, it has been suggested~\cite{Ladieu1996dti} that the conductivity jumps may be related to lifting of the Coulomb blockade by collective depinning. Transport in a 1D version of this model was recently studied in a paper co-authored by one of us~\cite{Fogler2006cba} but the case of extremely strong fields was not considered. It remains to be seen whether this model can yield a better agreement with the experiments.~\cite{Khavin1998slo}

It has been frequently speculated that the overheating is driven by the electron interactions, which we did not address here. The simplest way to introduce some interaction effects into the existing formalism is to consider larger dielectric constant $\epsilon > 1$. The importance of such effects requires further study.

Finally, as mentioned above, most of electron systems studied should behave as effectively 1D only at low enough $T$. The dimensional crossover as a function of temperature in a strip geometry has been studied by RR in Ref.~\onlinecite{Raikh1990sei}. It would be interesting to investigate the electric-field counterpart of this crossover.

In conclusion, we showed that numerical simulations such as those we carry out in this paper can serve as a valuable tool in studying VRH transport. We hope that our results would stimulate further experimental work on both ``conventional'' (semiconductor wires) and novel (nanotubes, nanofibers, graphene ribbons) 1D and quasi-1D materials.

This work is supported by the grant NSF DMR-0706654. We are grateful to
Colleague X, M.~E.~Raikh, A.~K.~Savchenko, and B.~I.~Shklovskii for valuable discussions and comments on the manuscript.


\end{document}